\documentclass[journal,onecolumn,12pt,twoside]{IEEEtranTCOM}

\usepackage[lmargin=1.0in, rmargin=1.0in,tmargin=1.0in,bmargin=1.0in]{geometry}
\usepackage{graphicx}
\usepackage[ruled,vlined]{algorithm2e}
\usepackage{epstopdf}
\usepackage{url}
\usepackage{algorithmic}
\usepackage{setspace}
\usepackage{subcaption}
\usepackage{caption}
\usepackage{epsfig}
\usepackage{bm}
\usepackage{slashbox}
\usepackage{latexsym}
\usepackage{amsmath}
\usepackage{amssymb}
\usepackage{epsfig}
\usepackage{amsthm}
\usepackage{fancyhdr}
\usepackage{cite}
\usepackage[T1]{fontenc}
\usepackage[utf8]{inputenc}
\usepackage{authblk}

\begin{document}

\newtheorem{theorem}{Theorem}
\newtheorem{proposition}{Proposition}
\newtheorem{lemma}{Lemma}
\newtheorem{example}{Example}
\newtheorem{notation}{Notation}
\newtheorem{definition}[theorem]{Definition}
\title{{Non-Cash Auction for Spectrum Trading in Cognitive Radio Networks: A Contract Theoretical Model with Joint Adverse Selection and Moral Hazard}}

\author{
Yanru Zhang,\thanks{Y. Zhang, M. Pan, and Z. Han are with the Department of Electrical and Computer Engineering, University of Houston, Houston, TX 77004 USA (e-mail: yzhang82@uh.edu; mpan2@uh.edu; zhan2@uh.edu).} \emph{Student Member, IEEE},
Lingyang Song,\thanks{L. Song is with the School of Electrical Engineering and Computer Science, Peking University, Beijing, China, 100871 (e-mail: lingyang.song@pku.edu.cn).} \emph{Senior Member, IEEE}, \\
Miao Pan, \emph{Member, IEEE},
Zaher Dawy,\thanks{Z. Dawy is with the Electrical and Computer Engineering Department, American University of Beirut, Riad El-Solh P.O. BOX 11-0236, Beirut 1107 2020, Lebanon (e-mail: zd03@aub.edu.lb).} \emph{Senior Member, IEEE},
and Zhu Han, \emph{Fellow, IEEE}
}\medskip

\maketitle

\begin{spacing}{1.0}
\begin{abstract}
In cognitive radio networks (CRNs), spectrum trading is an efficient way for secondary users (SUs) to achieve dynamic spectrum access and to bring economic benefits for the primary users (PUs). Existing methods requires full payment from SU, which blocked many potential ``buyers'', and thus limited the PU's expected income. To better improve PUs' revenue from spectrum trading in a CRN, we introduce a financing contract, which is similar to a sealed non-cash auction that allows SU to do a financing. Unlike previous mechanism designs in CRN, the financing contract allows the SU to only pay part of the total amount when the contract is signed, known as the down payment. Then, after the spectrum is released and utilized, the SU pays the rest of payment, known as the installment payment, from the revenue generated by utilizing the spectrum. The way the financing contract carries out and the sealed non-cash auction works similarly. Thus, contract theory is employed here as the mathematical framework to solve the non-cash auction problem and form mutually beneficial relationships between PUs and SUs. As the PU may not have the full acknowledgement of the SU's financial status, nor the SU's capability in making revenue, the problems of \emph{adverse selection} and \emph{moral hazard} arise in the two scenarios, respectively. Therefore, a joint \emph{adverse selection} and \emph{moral hazard} model is considered here. In particular, we present three situations when either or both \emph{adverse selection} and \emph{moral hazard} are present during the trading. Furthermore, both discrete and continuous models are provided in this paper. Through extensive simulations, we show that the \emph{adverse selection} and \emph{moral hazard} cases serve as the upper and lower bounds of the general case where both problems are present.
\end{abstract}
\end{spacing}

\newpage
\begin{spacing}{2.0}
\section{Introduction}

The recent popularity of hand-held mobile devices, such as smartphones, enables the inter-connectivity among mobile users without the support of Internet infrastructure. With the wide usage of such applications, the data outburst leads to a booming growth of various wireless networks and a dramatic increase in the demand for radio spectrum \cite{Letaief2009}. However, we are currently in the exhaustion of available spectrum. Thus, cognitive radio (CR) has emerged as a new design paradigm as its opportunistic access to the vacant licensed frequency bands, which releases the spectrum from shackles of authorized licenses, and at the same time improves the spectrum utilization efficiency \cite{Kim2008}.

Cognitive radio networks (CRNs) are designed based on the concept of dynamic spectrum sharing where CR users can opportunistically access the licensed spectrum \cite{Hossain2009}. In a CRN, the primary users (PUs) are the licensed users to utilize the frequency band, while the secondary users (SUs) can only utilize those spectrum resources when the PUs are vacant. Whenever the PUs are back, the SUs must vacate the frequency band immediately to guarantee the PUs' quality of service (QoS) \cite{Zhu2007}. In other words, in a CRN, the PUs have higher priority to use the frequency bands than the SUs. The SU can be regarded as a radio which is capable of changing its transmitter parameters and transmitting/operating frequency based on its interaction with the environment \cite{BWCM04}.

In CRNs, the problems of spectrum sensing and resource allocation have been extensively studied in previous works such as \cite{WXXTLL10}. While in this work, we focus on the economic aspect of spectrum trading between the PU and SU, which achieves SU's dynamic spectrum accessing/sharing and creates more economically benefits for the PU. The idea of the market-driven structure has initiated the spectrum trading model in CRNs, and promoted a lot of interesting researches on the design of trading mechanisms. Through spectrum trading, PUs can sell/lease their vacant spectrum for monetary gains, and SUs can purchase/rent the available licensed spectrum if they are in need of radio resources to support their traffic demands \cite{Gao2011JSAC}.

\begin{figure}[t]
    \centering
    \includegraphics[width=0.75\columnwidth,height=0.2\textwidth]{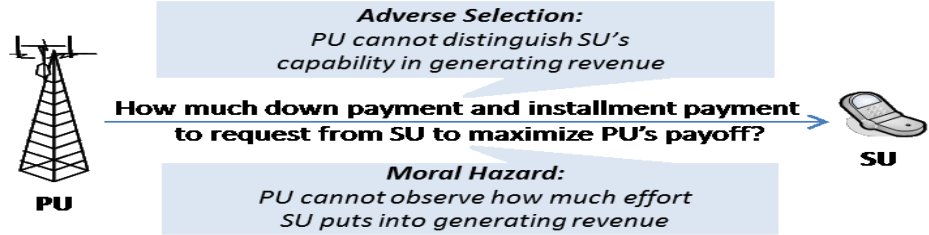}
    \caption{The problems of \emph{adverse selection} and \emph{moral hazard} in financing contract design.}
    \label{fig:model}
\end{figure}

There are many existing works addressing the spectrum trading problem by economic models such as game theory \cite{Han2011} and auction theory \cite{Zhou2009INFOCOM}, some recent works also adopted the \emph{adverse selection} model from contract theory \cite{Gao2011JSAC}. However, those works assume that full subscription fee is required before releasing the spectrum. The high entrance fee blocks many potential buyers, and thus limited the PU's expected revenue. In order to further improve the PU's expected revenue, we consider using a sealed non-cash auction that does not require full payment after the bid, but only a portion of it, and the rest comes from future revenues. Indeed, the non-cash auction has already been adopted in spectrum trading in the FCC $C$ block bandwidth auctions \cite{Landler1997}. The $C$ block was open only to small business, and bidders were allowed to pay only $10\%$ percent as the down payment and make installment payments from future revenues. The greatest advantage of non-cash auction over the traditional one-shot trading, and also the motivation for us to implement it here, is because it can raise higher revenue for seller. The reasons are, first, non-cash auction allows cash-constrained bidders to compete for the resources, i.e., an increase in the number of potential buyers. Second, in order to distinguish oneself from the other bidders, together with lower financial pressure, the bidders intend to bid higher than in cash auctions. Thus, the increases in the number of possible buyers and the value of bids together improve the seller's expected revenue.

Given this background, we have decided to take the advantages of non-cash auction to deal with the spectrum trading in CRN. Indeed, a sealed non-cash auction can be also seen as a financing contract, as we do for a house or a car \cite{Laffont1988}. That is, the SU only needs to pay part of the total amount at the point of signing the contract, known as the down payment. Then the spectrum can be released to the SU by the PU. Successively, the SU can utilize the spectrum to transmit package and generate revenue. Afterwards, the SU pays the rest of the loan, known as the installment payment. Thus, to solve the optimal bid in the non-cash auction, contract theory is adopted here to model and formulate the problem.

When the SU utilizes the spectrum to generate revenue, however, the PU may not have the knowledge of the SU's capability, i.e., what is the SU's probability of successful making profit, in which case the problem of \emph{adverse selection} arises \cite{Akerlof1995}. Moreover, the PU neither knows how much effort the SU exerts, where the problem of \emph{moral hazard} arises \cite{Roland.2000}. Such a combination of asymmetric information, \emph{adverse selection}, and \emph{moral hazard} makes the PU difficult to design an optimal contract, since the PU must consider the SU's current and future financial status \cite{scott2014financial}. Thus, to design such a financing contract, we must introduces a trade-off between inducing the SU to take the best action and extracting surplus. By referring to the contract theoretical model, we model the spectrum trading by a joint model which involves both \emph{adverse selection} and \emph{moral hazard} as shown in Fig. \ref{fig:model}. Later we will elaborate why the two special cases are not only serving as comparisons as contract theoretical models, but also comparisons between previous methods with the newly proposed one. As the \emph{adverse selection} only is the well known one-shot cash auction, and the \emph{moral hazard} only is the linear pricing strategy which is widely used in business and industry to pay wages and bonuses.

The main contributions of this paper are as follows.
\begin{itemize}
  \item A non-cash auction for spectrum trading is proposed, instead of the traditional one-shot cash auction. As far as we know, we are the first to adopt this model for spectrum trading in CRN.
  \item The innovative model of the financing contract that involves both \emph{adverse selection} and \emph{moral hazard} problems is brought out here to design the non-cash auction. In other words, we try to solve the optimal bid in the non-cash auction from the contract theory angle, instead of auction theory.
  \item The solutions to the problems under three different scenarios, i.e., the general case where both of the \emph{adverse selection} and \emph{moral hazard} are present, the two extreme cases where only \emph{adverse selection} or \emph{moral hazard} is present.
  \item Both the discrete and continuous models are provided, formulated, and discussed here. So that the models we have proposed can characterize more general scenarios, and be easily applied to other places.
  \item The analysis of how \emph{adverse selection} and \emph{moral hazard} affect the SU's activity and PU's contract design is provided, such as the proportions of down payment and installment payment, and how they affect the SU's incentives to generate revenues.
  \item Numerical results are provided to compare the optimal contracts under the three scenarios, as well as the previous methods.  The key parameters such as cost coefficient and revenue are studied to see their influences on the PU's and SU's payoffs.
\end{itemize}

The remainder of this paper is organized as follows. First, we will introduce the system model in Section \ref{sec:SystemModel}. Then, a literature review of spectrum trading and contract theory application in wireless networks are conducted in Section \ref{sec:related}. Next, the system model is described in Section \ref{sec:SystemModel}, and we formulated the PU's payoff maximization problems under the three scenarios in Section \ref{sec:ProbForm}. The continuous type model is then provided in Section \ref{sec:continuous}. Furthermore, we give analysis and illustrations of the solutions to the three optimization problems in Section \ref{sec:analysis}. The performance evaluation is conducted in Section \ref{sec:Simulations}. Finally, Section \ref{sec:Conclusion} draws the conclusion.


\section{Related Works}\label{sec:related}

Spectrum trading in CRNs has been extensively studied by using game theory \cite{Han2011}. Different game theoretical models have been adopted, such as potential game \cite{Nie2005DySPAN}, evolutionary game, non-cooperative game \cite{Niyato2009TMC}, and Stackelberg game \cite{Xie2012ICC}. Despite game theory, auction theory is another popular method to solve the spectrum trading problem. Different auction mechanisms such as double auction \cite{Zhou2009INFOCOM}. The fundamental one-shot auction has been extended to a real-time fast auction algorithms by \cite{Gandhi2007DySPAN}, and performance related auction by \cite{Wang2010TSMC}. Although one-shot auction based spectrum trading has been extensively studied, there is few work that has tangled the non-cash auction.

Contract theory has recently emerged into spectrum trading by some peer works. As far as we know, contract theory is first used to solve the problem of spectrum sharing in cognitive radio network (CRN) by \cite{Gao2011JSAC}. In this work, a primary user (PU) acts as an seller who sets the spectrum trading contract as \emph{(qualities, prices)}, and the second users (SUs) act as an buyer to choose a contract to sign. Another application in CRNs can be found in \cite{Duan2014TMC}, in which the authors model the PU and SUs as employer and employees, respectively. Then designing the \emph{(performance, reward)} in contract as \emph{(relaying power, spectrum accessing time)}, so that SUs will be rewarded with certain spectrum assessing time if they satisfied the relaying power requirement of the PU.

However, all of the above contract theory works fall into the applications of \emph{adverse selection} only problems in wireless networks. Compared to the wide adoption of the \emph{adverse selection} problem, the \emph{moral hazard} problem has hardly been applied in wireless networks by now. As we see, the literature in contract theory applied wireless networks, either \emph{adverse selection} or \emph{moral hazard} is considered when modeling the problem. In practice, however, it is usually hard to decide which of the two problems is more important, i.e., to figure out if it is a \emph{moral hazard} problem or an \emph{adverse selection} problem \cite{Prescott1984Eco}. Indeed, most incentive problems in practice are the combinations of \emph{moral hazard} and \emph{adverse selection}, as the non-cash auction in spectrum trading we considered in this paper. Thus, what we have done in this work, using joint contract theoretical model with joint \emph{adverse selection} and \emph{moral hazard} to solve the non-cash auction design problem, is of great innovation and importance.

\section{System Model}\label{sec:SystemModel}

In the non-cash auction, there are multiple SUs bidding for the PU's spectrum resource. To use contract theory to solve the problem, we can consider the spectrum trading between one PU and one SU from a different angle. We are designing the problem as there are $n$ different types of SUs using the sealed bid to bid for the spectrum. There are multiple types of SUs, each under a certain transmission scenario and has its own bid, which is their private information, to bid for the spectrum. Thus, we are designing the $n$ different bids (financing contracts) intended for each type of SU. While in this spectrum trading model, there is only one SU participate in the PU's auction. Thus, the SU will be the only winner and the corresponding financing contract (bid) will be signed, between one PU and one SU. The PU's spectrum is vacant, and the PU cannot generate any revenue from the vacant spectrum unless selling/leasing to the SU. Thus the spectrum trading is at no cost of the PU. We will fist give the definition of the financing contract. Then, the problems of \emph{adverse selection} and \emph{moral hazard} are discussed subsequently. Last, the payoff functions of both PU and SU, and social welfare are defined.

\subsection{Financing Contract}\label{subsec:contract}
The PU offers a financing contract $(t,r)$ to the SU for purchasing or leasing the spectrum, where $t$ is a down payment, and $r$ is an installment payment to be paid from the future revenue generated. The problem that the PU needs to solve is to find the optimal contract that can maximize its expected payoff from the spectrum trading by deciding how much down payment and installment payment to request from the SU.

The SU makes use of the spectrum to run its own ``business'', i.e., transmit data package to satisfy its own traffic need. Assuming that the transmission can only result in a success (data rate at the receiver satisfies the minimum threshold, $R_i\geq R_T$) or failure (data rate lower than the minimum threshold, $R_i< R_T$). Assuming that the SU can only transmit data package when PU is vacant and the spectrum assigned to each SU is orthogonal. Thus, we consider neither the interference between the PU and SU, nor the interference among different SUs. Furthermore, the channel quality offered by the PU is the same for all types of SU, thus no difference between channel gains. Given by the Shannon-Hartley theorem, the transmission data rate in the CRN as
\begin{align}\label{eq:rate}
R_i=W\log_2\left(1+\frac{p_i|h_{i}|^n}{d_iN_0}\right),
\end{align}
where $W$ is the channel bandwidth, $p_i$ is the SU's transmit power which can be adjusted during the transmission, $h_{i}$ is the channel gain, $d_i$ is the transmission distance which is fixed during transmission, and $N_0$ is the additive white Gaussian noise (AWGN). Hereinafter, without loss of generality, we assume that $W=1$, the channel condition and AWGN are identical for every SU.

We have no idea what the threshold is, but that the higher the achieved data rate, the higher the probability of a successful transmission. If the data package is successfully transmitted, the SU will receive a revenue of $R\geq r \geq 0$, otherwise, it receives a revenue of $0$, i.e., the revenue realizations of an SU is $X \in \{0, R\}$. We assume that there is no installment payment if the SU cannot generate revenue from utilizing the spectrum, i.e., $r=0$ if $X=0$, and $r>0$ only when $X=R$.

\subsection{Problem of Adverse Selection}\label{subsec:adverse}
As we mentioned in the introduction part, the PU is unaware of the SU's capability of successfully running its own ``business'', i.e., successfully transmit the data package. Learning from (\ref{eq:rate}), we see that under the same transmission power, the shorter the transmission distance, the higher the data rate and thus higher probability to transmit package successfully, vice versa. Since the transmission distance $d_i$ is fixed, we can define the SU's capability type $\theta$ as a variable related to the inverse of the transmission distance $\frac{1}{d}$  between secondary transmitter and receiver. Since the transmission distance is a continuous variable, the type of SU can be a discrete set or a continuous region. In this work, we will mainly focus on the discrete type case, and briefly provide the continuous type model in the later part.

Assuming that the PU is aware of a part of the information such as the total number of SU types and the value of each type (range of transmission distance), we have the following definition of SU type.
\begin{definition}
  The SU's capability belongs to $n$ different types $\theta \in \{\theta_1, \ldots, \theta_i, \ldots,\theta_n\}$ and satisfies the following inequality
  \begin{equation}
  \theta_1 < \cdots < \theta_i < \cdots < \theta_n, \quad i \in \{1, \cdots, n\},
  \end{equation}
  which means less or more able at successfully transmitting a data package, and thus generating revenue.
\end{definition}

From a marketing perspective, a seller lowers its down payment will attract more potential customers with limited cash available, and increases its expected income. However, without background check of those financing customers, the seller also bears the risk of unable to collect the loan back. Thus, there is a tradeoff between attracting more customers and money back guarantee. Similarly, in the CRN spectrum trading, giving the same transmission power, the shorter the transmission distance, the more capable an SU can successfully transmit the data package at a required data rate, and thus pay off the installment payment.

The PU has the information that in total there are $n$ types of SUs in the CRN, but does not know which certain type an SU belongs to, and this is where the problem of \emph{adverse selection} arises. However, the PU has a priori that an SU is type $\theta_i$ with probability $\beta_i \in [0,1]$, and
\begin{equation}
  \sum_{i=1}^n \beta_i=1.
  \end{equation}
To overcome this problem of \emph{adverse selection}, the PU offers a list of contracts to different types of SU as $(t_i,r_i)$, $i\in \{1,\ldots,n\}$ for different types of SUs to select.

The higher the type of SU, the more likely it can successfully transmit the data package and receive the revenue. Thus, the PU can lower the down payment, and increase the installment payment to attract more SU to buy spectrum. On the contrary, the PU will request higher down payment if it believes an SU belongs to a lower type and is less likely to pay the installment payment in the future. So that the PU can find a balance between attracting more customers and collecting loan back.

\subsection{Problem of Moral Hazard}\label{subsec:moral}

To successfully transmit a data package, higher transmission power will increase the probability of achieving the minimum data rate at the receiver under the same transmission distance. Thus, regarding  the effort $e$ it exerted to make this ``business'' a success as a variable positively related to the transmission power of the SU. The SU's operation cost $\psi$ on the spectrum is a convex function of effort $e$, which is
\begin{equation}
\psi(e)= \frac{c}{2}e^2,
\end{equation}
where $c$ is the cost coefficient. We denote the efforts exerted by different type of SUs as $e_i$, $i\in \{1,\ldots,n\}$. The operation cost here is defined as a square function of power, but not restricted to that. The exact form of the cost function can be adjusted to different forms when applied to different scenarios.

The PU has no idea of what transmission power the SU has, and this is where the problem of \emph{moral hazard} arises. To overcome this problem of \emph{moral hazard}, the PU must carefully design the amount of down payment and installment payment in each contract $(t_i,r_i)$, $i\in \{1,\ldots,n\}$, so that the SU can select the optimal amount of effort to exert and PU's payoff can be maximized.
\subsection{Mixure of Adverse Selection and Moral Hazard}\label{subsec:mixed}

As we defined in the previous two subsections, the part $p_i/d_i$ in the data rate function determines the probability of a successful transmission. Thus, we can define $\theta_ie \in (0,1)$ as the probability of a successful transmission and make revenue $R$, through normalization of $1/d$ and $p_i$. Since the inverse of transmission distance $1/d$ is positively related to the type of SU $\theta_i$, i.e., its capability of successful transmitting a data package under the same transmission power, and the transmission power $p_i$ is positively related to the effort an SU exerted $e$.

\subsection{Payoff of SU}\label{subsec:SU}
Based on our assumption that, the instalment payment will only be made when the business is a success. Otherwise, $0$ payment will be made. The revenue $R$ minus the installment payment $r_i$ is the SU's income. Thus, the SU's expected revenue is
\begin{equation}
\theta_i e_i(R-r_i)+ (1-\theta_i e_i)0=\theta_i e_i(R-r_i),  \quad i\in\{1,\ldots,n\}.
\end{equation}

The expected payoff of the SU with capability $\theta_i$ under contract $(t_i,r_i)$ takes the form of
\begin{equation}
U_{SU_i} =\theta_i e_i(R-r_i)-t_i-\frac{c}{2}e_i^2,  \quad i\in\{1,\ldots,n\}.
\end{equation}
The SU's expected payoff is thus the expected income minus the down payment and the cost of operation.

\subsection{Payoff of PU}\label{subsec:PU}
Similarly, the payoff of the PU trading with $\theta_i$ is
\begin{equation}
U_{PU_i} =t_i+\theta_ie_ir_i,  \quad i\in\{1,\ldots,n\},
\end{equation}
which is the summation of the down payment and expected installment payment.

Under the problem of \emph{adverse selection}, the PU only knows the probability of an SU belongs to a certain type $\theta_i$. Thus, we define the expected payoff of the PU as
\begin{align}
U_{PU} =\sum_{i=1}^n \beta_i(t_i+\theta_ie_ir_i), \quad i\in \{1,\ldots,n\}.
\end{align}
The PU's expected payoff is summation of the SU's expected payoff in each type.

\subsection{Social Welfare}\label{subsec:social}
The social welfare is defined as the summation of the expected payoff of both PU and SU as
\begin{align}
U &=\sum_{i=1}^n (U_{PU_i}+U_{SU_i}) \\ \nonumber
  &=\sum_{i=1}^n \beta_i(\theta_i e_iR-\frac{c}{2}e_i^2),\\
  & \quad i\in \{1,\ldots,n\}. \nonumber
\end{align}
The social welfare is the expected revenue from the spectrum trading minus the SU's operation cost during the data transmission process, and down payment and installment payment items are canceled out.

\section{Problem Formulation}\label{sec:ProbForm}

In this section, we will solve the PU's problem by considering three scenarios, i.e., the general case where both \emph{moral hazard} and \emph{adverse selection} are present, the two extreme cases where only \emph{moral hazard} or \emph{adverse selection} is present, respectively.

\subsection{Optimal Contract with Both Adverse Selection and Moral Hazard}\label{subsec:both}
The PU's payoff maximization problem is formulated as
\begin{align}\label{eq:Opt1}
& \max_{(t_i,r_i)}\sum_{i=1}^n \beta_i(t_i+\theta_ie_ir_i),
\\
& \quad \quad s.t. \nonumber
\\
&(IC)\quad \theta_i e_i(R-r_i)-t_i-\frac{c}{2}e_i^2 \geq\theta_i e_i'(R-r_j)-t_j-\frac{c}{2}e_i'^2, \nonumber
\\
&(IR) \quad \theta_i e_i(R-r_i)-t_i-\frac{c}{2}e_i^2 \geq 0,  \nonumber
\\
& \quad \quad \forall j\neq i, \quad  i,j \in\{1,\ldots,n\},\nonumber
\end{align}
where $e_i'$ is the effort of $\theta_i$ SU when selecting contract $(t_j, r_j)$. The IC constraint stands for incentive compatibility, which means the SU can only maximize its expected payoff by selecting the financing contract that fits its own capability. The IR constraint stands for individual rationality, which provides the SU necessary incentives to sign the contract.

Taking the first derivative of SU's expected payoff with respect to effort $e$, we have the SU's optimal choice of effort:
\begin{equation}
e_i^*=\frac{1}{c}\theta_i (R-r_i), \quad i\in \{1,\ldots,n\}.
\end{equation}
Similarly, we have $e_i'^*=\frac{1}{c}\theta_i (R-r_j)$. As we can see, the SU's optimal choice of effort $e_i^*$ is independent of $t_i$ but is decreasing in $r_i$. In other words, the SU will have fewer incentives to exert more effort, if it must share more of the generated revenue with the PU, regardless of the amount of the down payment $t_i$. The decrease of effort $e$ directly affects the probability of successfully generating revenue $R$. Thus, it is critical to balance the tradeoff between providing necessary incentives and request more installment payment.

Replacing SU's choice of effort $e_i$ and $e_i'$ in (\ref{eq:Opt1}), we have
\begin{align}\label{eq:Opt12}
& \max_{(t_i,r_i)} \sum_{i=1}^n \beta_i(t_i+\theta_ie_ir_i),
\\
s.t. \quad & (IC) \quad\!\! \frac{1}{2c}[\theta_i(R-r_i)]^2-t_i\geq \frac{1}{2c}[\theta_i(R-r_j)]^2-t_j, \nonumber
\\
& (IR) \quad\!\! \frac{1}{2c}[\theta_i(R-r_i)]^2-t_i\geq 0, \nonumber
\\
& \quad \quad \forall j\neq i, \quad  i,j \in\{1,\ldots,n\}.\nonumber
\end{align}

In this problem, it is not possible to decide a priority which of the two incentive problems is the more important, i.e., to disentangle the \emph{moral hazard} from the \emph{adverse selection} dimension. In the following section, we will detail the respective roles of \emph{moral hazard} and \emph{adverse selection}, and show the implications of their simultaneous presence.

Solving the problem of the PU can be done by relying on the pure \emph{adverse selection} methodology detailed in \cite{Bolton.2004}. Specifically, the analysis shows that only the IR constraint of the $\theta_1$ SU and the IC constraint between $\theta_i$ and $\theta_{i-1}$, which is called local downward IC (LDIC) constraint will be binding. Therefore, the PU has to solve
\begin{align}\label{eq:Opt13}
& \max_{(t_i,r_i)}  \sum_{i=1}^n \beta_i[t_i+ \frac{1}{c}\theta_i^2(R-r_i)r_i],
\\
& \quad \quad s.t. \nonumber
\\
& (IC) \quad\!\! \frac{1}{2c}[\theta_i(R-r_i)]^2-t_i = \frac{1}{2c}[\theta_i(R-r_{i-1})]^2-t_{i-1}, \nonumber
\\
& (IR) \quad\!\! \frac{1}{2c}[\theta_1(R-r_1)]^2-t_1 = 0. \nonumber
\\
&  \quad  i\in\{2,\ldots,n\}.\nonumber
\end{align}

To solve this optimization problem by using the Lagrangian multiplier method, assuming $\mu_i$ and $\nu$ are the Lagrangian multipliers for the IC and IR constraints. First, based on the optimization problem, we have the Lagrangian as
\begin{align}
\mathcal{L}&=\sum_{i=1}^n \{\beta_i[t_i+ \frac{1}{c}\theta_i^2(R-r_i)r_i] \}\nonumber \\
&+ \mu_i \{\frac{1}{2c}[\theta_i(R-r_i)]^2-\frac{1}{2c}[\theta_i(R-r_{i-1})]^2 +t_{i-1}-t_i \}\nonumber \\
&+ \nu \{\frac{1}{2c}[\theta_1(R-r_1)]^2-t_1\}.
\end{align}
The partial derivatives regarding $t_i$ and $r_i$ when $i=n$ are
\begin{align}
\frac{\partial \mathcal{L}}{\partial t_n}=0 &\Leftrightarrow \beta_n r_n=(1-\mu_n) (R-r_n), \\
\frac{\partial \mathcal{L}}{\partial r_n}=0 &\Leftrightarrow \beta_n = \mu_n.
\end{align}
For $i\in \{1,\ldots, n-1\}$, we have
\begin{align}
\frac{\partial \mathcal{L}}{\partial t_i}=0 &\Leftrightarrow  \beta_i\theta_i^2(R-2r_i)=(\mu_i \theta_i^2 - \mu_{i+1} \theta_{i+1}^2)(R-r_i), \\
\frac{\partial \mathcal{L}}{\partial r_i}=0 &\Leftrightarrow  \beta_i = \mu_i -\mu_{i+1}.
\end{align}

With the $2n$ equations, we can solve $r_n=0$ and $r_i$ for $i\in \{1,\ldots, n-1\}$ as
\begin{align}
r_i=\frac{(\beta_i\theta_i^2-\mu_i \theta_i^2 + \mu_{i+1} \theta_{i+1}^2)R}{\mu_{i+1} \theta_{i+1}^2-\mu_i \theta_i^2 +2\beta_i\theta_i^2},
\end{align}
with $\mu_i = \beta_i+\mu_{i+1}$ for $i\in \{1,\ldots, n-1\}$ and $\mu_n=\beta_n$. Thus, we can further simplify $r_i$ as
\begin{align}
r_i=\frac{\mu_{i+1}(\theta_{i+1}^2-\theta_i^2)R}{\mu_{i+1} (\theta_{i+1}^2-\theta_i^2)+\beta_i\theta_i^2}.
\end{align}

Taking the solution of $r_i$ into the IR and IC constraints of (\ref{eq:Opt13}), the solutions to the down payment can be obtained as
\begin{align}
t_1 =\frac{1}{2c}[\theta_1(R-r_1)]^2.
\end{align}
Then, backward deduction from the IC constraint, we have
\begin{align}
t_i=\frac{1}{2c}[\theta_i(R-r_i)]^2-\frac{1}{2c}[\theta_i(R-r_{i-1})]^2 +t_{i-1},
\end{align}
for $i\in\{2,\dots, n\}$.

The optimal contract is thus such that the high capable SU achieves the optimal effort efficiency. But there is an effort distortion for the lower capable SU. The extent of the distortion between high capability and low capability SUs depends on the size of the capability differential $(\theta_i^2-\theta_{i-1}^2)$ and on the PU's prior $\beta$: The more confident the PU is that it faces a high capable SU, the larger is the SU's installment payment $r$ and the less down payment $t$.
\subsection{Optimal Contract with Adverse Selection Only}\label{subsec:AdverseOnly}
When the SU's effort is observable, but capability is unavailable to the PU, the \emph{moral hazard} problem is removed, but only \emph{adverse selection} present. In this case, the PU is optimal to treat each SU separately for different capabilities and the problem reduces to
\begin{align}\label{eq:Opt2}
& \max_{(t_i,r_i)} t_i+ \frac{1}{c}\theta_i^2(R-r_i)r_i,
\\
s.t. \quad & (IR) \quad\!\! \frac{1}{2c}[\theta_i(R-r_i)]^2-t_i\geq 0,
\quad i \in\{1,\ldots,n\}.\nonumber
\end{align}
Since the IR constraint is binding, the problem becomes
\begin{equation}
\max_{r_i}  \frac{1}{2c}[\theta_i(R-r_i)]^2+ \frac{1}{c}\theta_i^2(R-r_i)r_i.
\end{equation}
The solution for this maximization problem is
\begin{equation}
t_i =\frac{1}{2c}\theta_i^2R^2,
\end{equation}
\begin{equation}
r_i =0.
\end{equation}
When there is only \emph{adverse selection} present, it is optimal for the PU to sell the spectrum for cash only, and not keep any financing participation in, i.e., all money paid in the down payment, no installment payment requested. Since the PU does not want to place any risk on if the loan can be collected back.

\subsection{Optimal Contract with Moral Hazard Only}\label{subsec:MoralOnly}
Suppose that the PU is able to observe the SU's capability but cannot observe SU's effort, then the only remaining incentive problem is \emph{moral hazard}. To avoid the \emph{moral hazard}, now it is optimal to fix the each SU's effort level at a certain level $\widehat{e}$ which is consistent with its capability. The PU's problem then reduced to
\begin{align}\label{eq:Opt3}
& \max_{(t_i,r_i)} \sum_{i=1}^n \beta_i(t_i+\theta_ie_ir_i),
\\
s.t. \quad & (IC) \quad\!\! \theta_i\widehat{e}(R-r_i)-t_i\geq \theta_i\widehat{e}(R-r_j)-t_j, \nonumber
\\
& (IR) \quad\!\! \theta_i\widehat{e}(R-r_i)-t_i- \frac{c}{2}\widehat{e}^2\geq 0, \nonumber
\\
& \quad \quad   \forall j\neq i,\quad  i,j \in\{1,\ldots,n\}. \nonumber
\end{align}
The solution for this problem is
\begin{equation}
t_i =-\frac{1}{2}c\widehat{e}^2 < 0,
\end{equation}
\begin{equation}
r_i = R.
\end{equation}
Intuitively, the down payment should be larger than or equal to $0$. However, in this case, the SU has a negative down payment, i.e., the PU has to pay $\frac{1}{2}c\widehat{e}^2$ to the SU, instead. This result is due to the fact that the PU pushes the SU as hard as it can and asks for $100$\% of the future revenue from the SU. So that the SU will exert the maximum effort to pay the loan.

The simplicity of the preceding solutions is due to the simplified setup. However, neither extreme formulation is an adequate representation of the basic problem in practice, nor that it is necessary to allow for both types of incentive problems to have a plausible description of the spectrum trading \cite{darrough1986moral}. As we see from the general case, the optimal menu of contracts where both types of incentive problems are present is a combination of the two extreme solutions that we have highlighted.


\section{Extension to Continuous Type}\label{sec:continuous}
In the previous case, there are $n$ type of SUs from $\theta_1$ to $\theta_n$. In practice, the number of SU types can be infinite. To extend the previous discrete model into a general one that can fir into more general cases, we can moderate the discrete type into a continuous one as follows.

\subsection{System Model}
First of all, now we have all SUs' types are independently and identically distributed (IID) and drawn from the distribution $f(\theta)$ over the interval $[\underline{\theta}, \overline{\theta}]$, with $F(\underline{\theta}) = 0$ and $F(\overline{\theta}) = 1$. The cumulative distribution function $F(\theta)$ is strictly increasing and differentiable over the interval.

The other definitions of revenue, down payment, installment payment, and effort are the same as in the case of discrete type. The revenue is still $R$ if the project is a success. While the forms of the down payment and installment payment, i.e., the contract has changed to the continuous forms as $(t(\theta), r(\theta))$, as well as the effort $e(\theta)$ put by type $\theta$ SU. As a result, the forms of the payoffs and social welfare are now need to be adjusted as follows.

\subsubsection{Payoff of SU}\label{subsubsec:SU}
The expected payoff of the SU with capability $\theta$ under contract $(t(\theta), r(\theta))$ is still the expected income (the revenue $R$ minus the installment payment $r(\theta)$) minus the down payment and operation cost. But the payoff has a continuous form now as follows.
\begin{equation}
U_{SU(\theta)} =\theta e(\theta)[R-r(\theta)]-t(\theta)-\frac{c}{2}e^2(\theta),  \quad \theta \in [\underline{\theta}, \overline{\theta}].
\end{equation}

\subsubsection{Payoff of PU}\label{subsubsec:PU}
Similarly, the payoff of the PU trading with $\theta$ SU is the summation of the down payment and expected installment payment with the continuous form as follows.
\begin{equation}
U_{PU(\theta)} =t(\theta)+\theta e(\theta)r(\theta),  \quad \theta \in [\underline{\theta}, \overline{\theta}],
\end{equation}

The PU's expected payoff is integration of the SU's expected payoff over all types.
\begin{align}
U_{PU} =\int_{\underline{\theta}}^{\overline{\theta}} f(\theta)[t(\theta)+\theta e(\theta) r(\theta)]d\theta, \quad \theta \in [\underline{\theta}, \overline{\theta}].
\end{align}

\subsubsection{Social Welfare}\label{subsubsec:social}
The social welfare is now the integration of the expected payoff of both PU and SU as
\begin{align}
U &=\int_{\underline{\theta}}^{\overline{\theta}} f(\theta) [U_{PU(\theta)}+U_{SU(\theta)}] \\ \nonumber
  &=\int_{\underline{\theta}}^{\overline{\theta}} f(\theta) [\theta e(\theta)R-\frac{c}{2}e^2(\theta)]d\theta,\\
  & \quad \theta \in [\underline{\theta}, \overline{\theta}]. \nonumber
\end{align}
The social welfare is still the expected revenue from the spectrum trading minus the SU's operation cost during the data transmission process, and down payment and installment payment items are canceled out.

\subsection{Problem Formulation}\label{subsec:ProbForm}

In this subsection, we rewrite the previous problem formulation into the continuous form.

\subsection{Optimal Contract with Both Adverse Selection and Moral Hazard}\label{subsubsec:both}
The PU's payoff maximization problem is formulated as
\begin{align}\label{eq:Opt1}
& \max_{(t(\theta),r(\theta))}\int_{\underline{\theta}}^{\overline{\theta}} f(\theta)[t(\theta)+\theta e(\theta) r(\theta)]d\theta,
\\
& \quad \quad s.t. \nonumber
\\
&(IC)\quad \theta e(\theta)[R-r(\theta)]-t(\theta)-\frac{c}{2}e^2(\theta)  \geq \theta e'(\theta)[R-r(\theta)]-t(\theta)-\frac{c}{2}e'^2(\theta), \nonumber
\\
&(IR) \quad \theta e(\theta)[R-r(\theta)]-t(\theta)-\frac{c}{2}e^2(\theta) \geq 0,  \nonumber
\\
& \quad \quad \theta \in [\underline{\theta}, \overline{\theta}],\nonumber
\end{align}
where $e'(\theta)$ is the effort exerted by type $\theta$ SU when selecting the other type of contract $(t(\theta'), r(\theta'))$ which is not suitable for its own type.

The way of solve this continuous type case is similar to the discrete type case. That is, reduce the IC and IR constraints first, they solve the reduced constraints problem. The steps to reduce the IC and IR constraints can be found in our previous work \cite{Zhang2015JSAC}. The detailed steps to solve the reduced constraint problem can also be found at our previous work \cite{Li2016ICCC}.

\section{Discussion}\label{sec:analysis}
In this section, we will give a discussion of the solutions of the previous optimization problems, to see how \emph{adverse selection} and \emph{moral hazard} affect the SU's selection of effort, and lead to the change in PU's payoff.
\subsection{Affects of Adverse Selection}\label{subsec:Anladverse}

With the presence of \emph{adverse selection}, SUs benefit from the information asymmetry since they can pretend to be high capable and pay lower amount of down payment at the signing of the contract. Thus, the PU tries to extract those information so that to avoid the situation of unable to receive the installment payment. However, subject to the IC constraint in the optimization problem, the SU can only achieve the maximum payoff when selecting the type of contract in consistent with its own type. Thus, the SU has the incentive to choose the right contract, and automatically reveals its true type. In other words, the contract is designed in a way such that the SU is forced to tell its true type of capability.

In the special case when only \emph{adverse selection} presents, the result is indeed an optimal bid in a cash auction. That is, multiple types of SUs with their own private information about their finical status and capability come and bid the spectrum with the maximum amount of money they can afford. Thus, we see that in the \emph{adverse selection} only case, there is no installment payment in the financing contract, but only down payment. It has been applied into spectrum trading in CRN from the contract theoretical aspect by \cite{Duan2014TMC}.

\subsection{Affects of Moral Hazard}\label{subsec:Anlmoral}

When \emph{moral hazard} presents, the PU is unable to know the SU's effort (transmission power). Given the minimum data rate $R_T$, as long as the data rate at receiver $R_i$ is equal to or larger than $R_T$, the SU can receive the revenue, no matter what level of effort (transmission power) it has, neither the PU will know about it. Thus, the SU has the incentive to adjust its transmission power to the most efficiency level to lower its operation cost \cite{Laffont1993}. For example, in the case of short transmission distance, the transmission power can be reduced while still guarantee the data rate at the receiver. So with a lower operation cost, the SU can increase its own payoff. However, in the opposite case, to achieve the same data rate at the receiver, the SU must increase the transmission power, and thus encounters a larger operation cost. In summary, at the presence of \emph{moral hazard}, the SU has the incentive to reduce its cost, but also takes the burden of more risks.

On the contrary, when the PU has the perfect information about the amount of effort the SU exerted, as the case in Section \ref{subsec:AdverseOnly}, any cost saving behavior will be monitored by the PU, and the contract will be redesigned to extract those savings from the SU to the PU. Thus, the SU will have no incentive for cost reduction, for any savings it has made does not belong to itself. Therefore, the presence of \emph{moral hazard} provides incentives for the SU to reduce the operation cost, but \emph{adverse selection} cannot.

Indeed, the \emph{moral hazard} only is the pay-after-work contract which is widely used in business and industry to for wages and bonuses, similar to the piece rate mechanism which pay the money afterwards. Under this incentive mechanism, the employers have the motivation to work as hard as possible. Since their payoffs are positively correlated with the amount of effort they have exerted.


\section{Simulation Results}\label{sec:Simulations}

In this section, we first give an analysis about the structure of the financing contract when both \emph{adverse selection} and \emph{moral hazard} are present by varying the parameters such as the cost coefficient, revenue, and the SU's probability of being the highest type SU $\theta_i$, when $i=n$. Then, we conduct comparisons among the PU's and SU's payoffs, and social welfare among the three scenarios we have proposed. As we mentioned in the previous section, the two extreme cases where only \emph{adverse selection} and \emph{moral hazard} are also representations of the existed works: one-shot cash auction and linear pricing, respectively. Thus, the comparisons we present are not only the comparisons among the three mechanisms that we have proposed, but also companions between the newly proposed non-cash auction and previous works.

For the simulation set up, we consider $10$ different types of SUs in the system, i.e., $n=10$. Further more, each type of SU, the value of type is set up as $\theta_i=i$ for simplicity, so that the monotonicity constraint in Definition 1 is satisfied, i.e., the higher the type (capability), the higher the value of $\theta$. In addition, the distribution of different types of SUs is assumed as a uniform distribution, i.e., $\beta_i=\frac{1}{n}$. For the other parameters such as cost coefficient $c$, revenue $R$, probability of being a high type SU $\beta$, will be varied to see how they affect the payoffs of the PU and SU, as well as the social welfare.

\subsection{Financing Contract Analysis}\label{subsec:Analy}

To analysis the structure of the financing contract, we are going to show the contract for three types of SUs. There are two special cases, one is the highest type with $i=n$ which is denoted as $\theta_H$ in the simulations, the other is the lowest type $\theta_L$ of SU $i=1$. The reason for this selection is due to the speciality of the $\theta_H$ and $\theta_L$ SUs. This result is not only given in Section. \ref{subsec:both}, but also by previous works such as \cite{Gao2011JSAC} and \cite{Zhang2015JSAC}. For the rest types of SUs, the contract they receive have similar properties. Thus we only pick one of them for analysis, and the third one is randomly picked up among $i=1, \ldots, n-1$ which is denoted as $\theta_M$ in the simulations. When studying the effect of one parameter, the other parameters will be fixed as follows. When varying the cost coefficient, we fix the revenue $R=0.5$, type $i$ SU's probability $\beta=0.1$. While we fix the cost coefficient $c=5$ when varying the revenue $R$. When studying the probability $\beta$ of facing a high type SU, the solution for general cases requires quite amount of additional works which can not be covered in this paper due to content limitation. Thus we study the $n=2$ case only, while fixing the cost coefficient $c=4$ revenue $R=0.5$.

\begin{figure*}
 \begin{subfigure}[b]{0.32\textwidth}
 \centering
 \includegraphics[width=\columnwidth,height=0.9\textwidth]{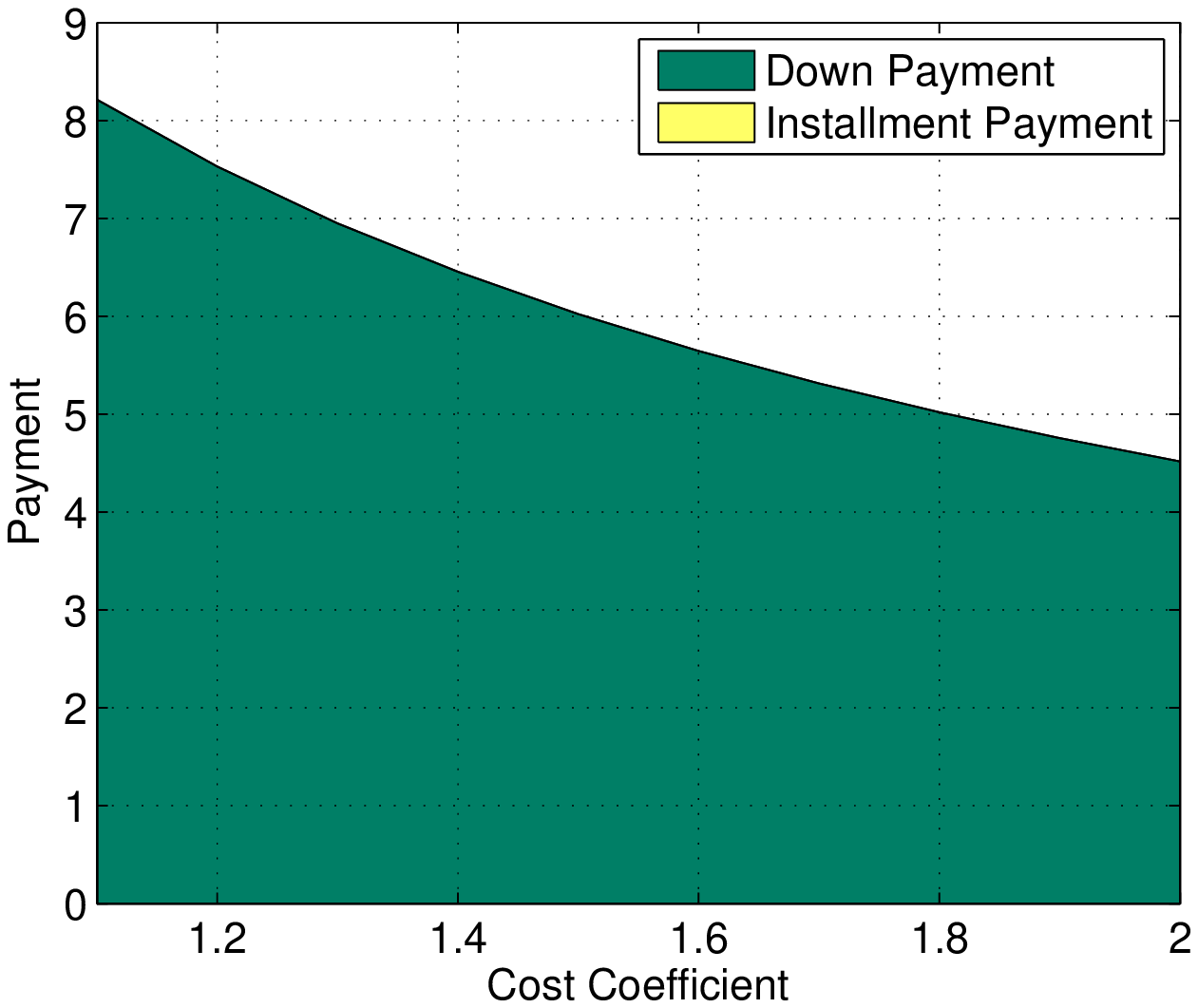}
 \caption{Cost Coefficient $c$}
 \label{fig:Highc}
 \end{subfigure}
 \begin{subfigure}[b]{0.32\textwidth}
 \centering
 \includegraphics[width=\columnwidth,height=0.9\textwidth]{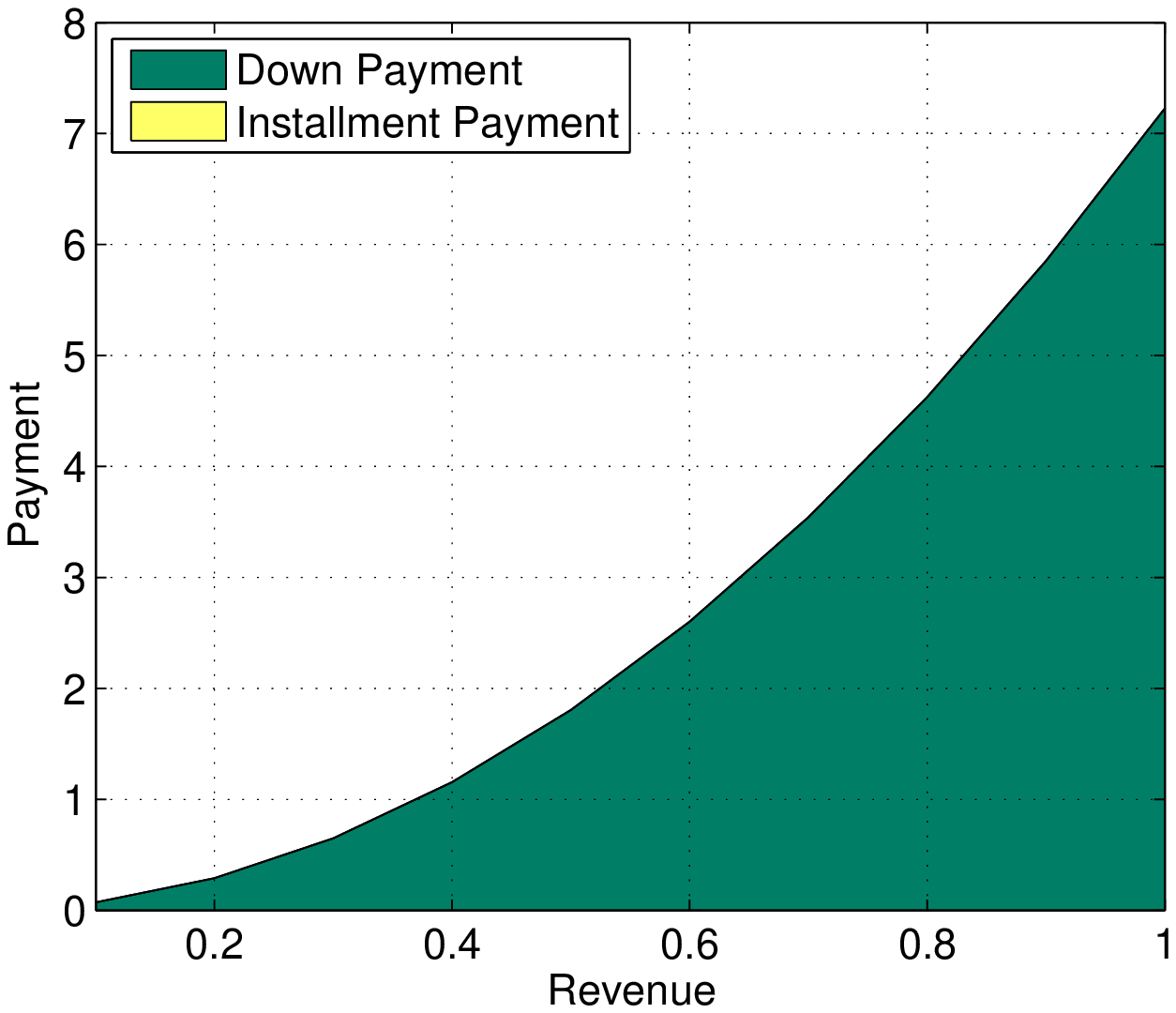}
 \caption{Revenue $R$}
 \label{fig:HighR}
 \end{subfigure}
  \begin{subfigure}[b]{0.32\textwidth}
 \centering
 \includegraphics[width=\columnwidth,height=0.9\textwidth]{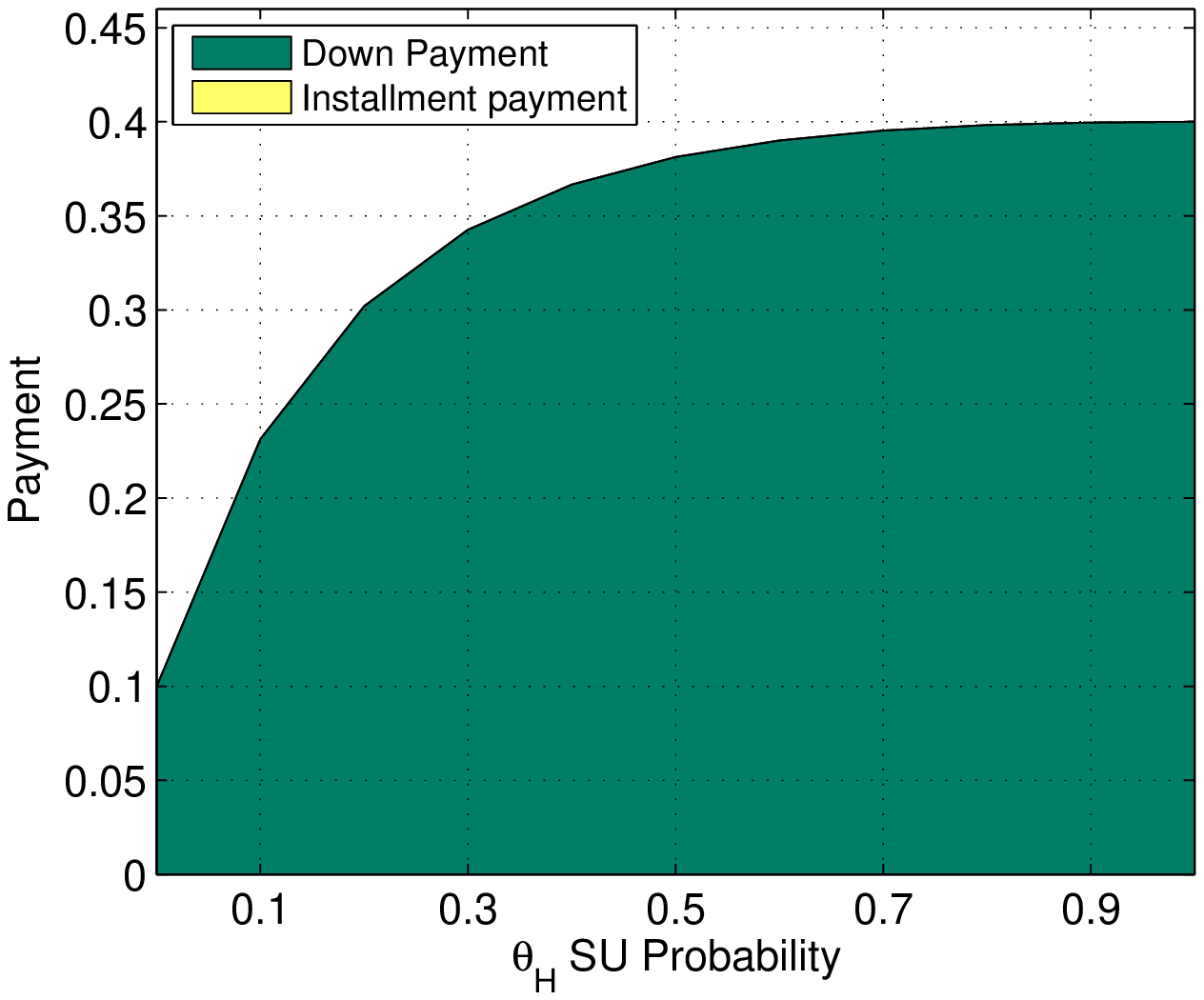}
 \caption{$\theta_H$ SU Probability $\beta$}
 \label{fig:Highb}
 \end{subfigure}
\centering
\caption{The financing contract for $\theta_H$ SU as parameters vary.}
\label{fig:High}
\end{figure*}

\begin{figure*}
 \begin{subfigure}[b]{0.32\textwidth}
 \centering
 \includegraphics[width=\columnwidth,height=0.9\textwidth]{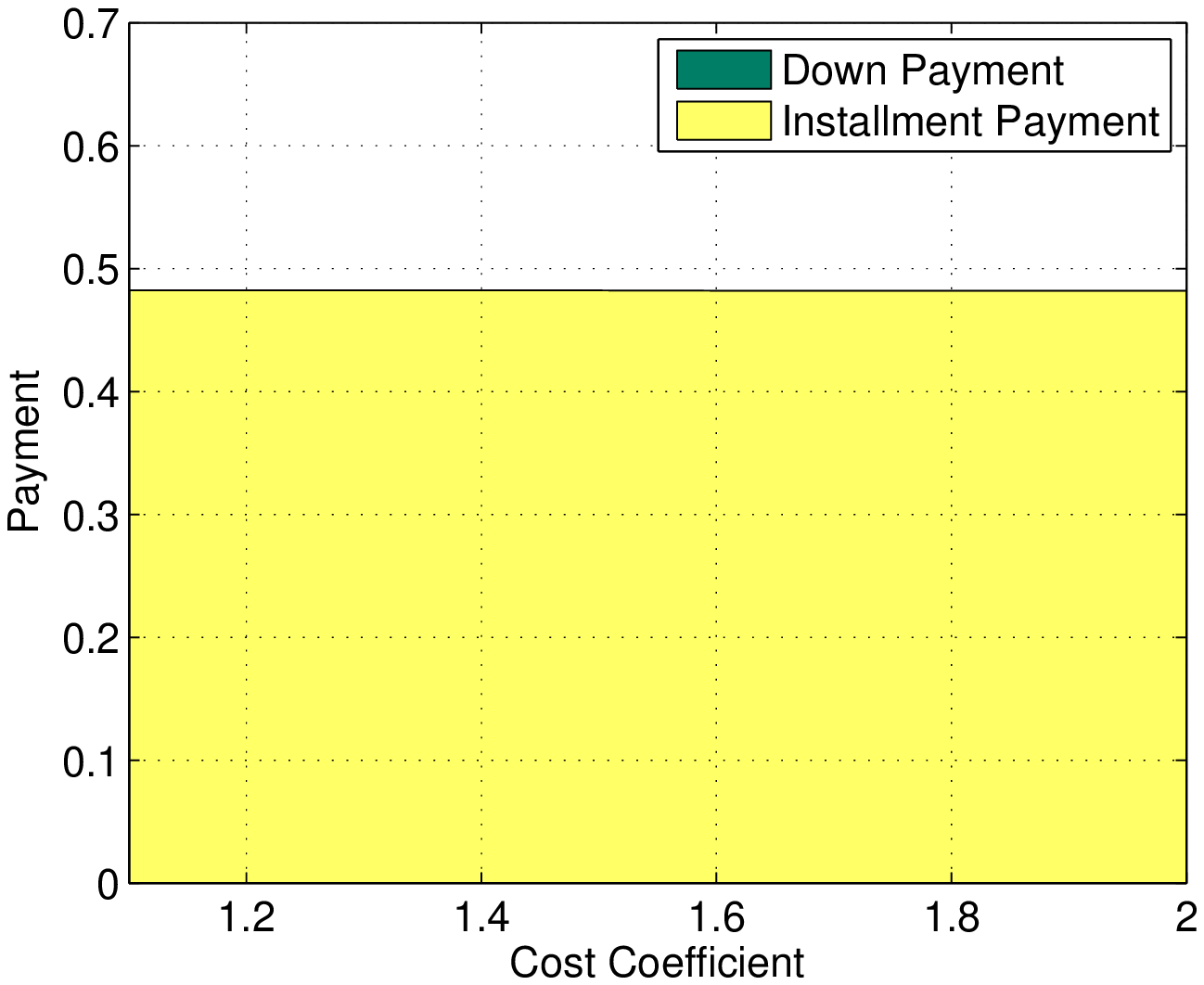}
 \caption{Cost Coefficient $c$}
 \label{fig:Lowc}
 \end{subfigure}
 \begin{subfigure}[b]{0.32\textwidth}
 \centering
 \includegraphics[width=\columnwidth,height=0.9\textwidth]{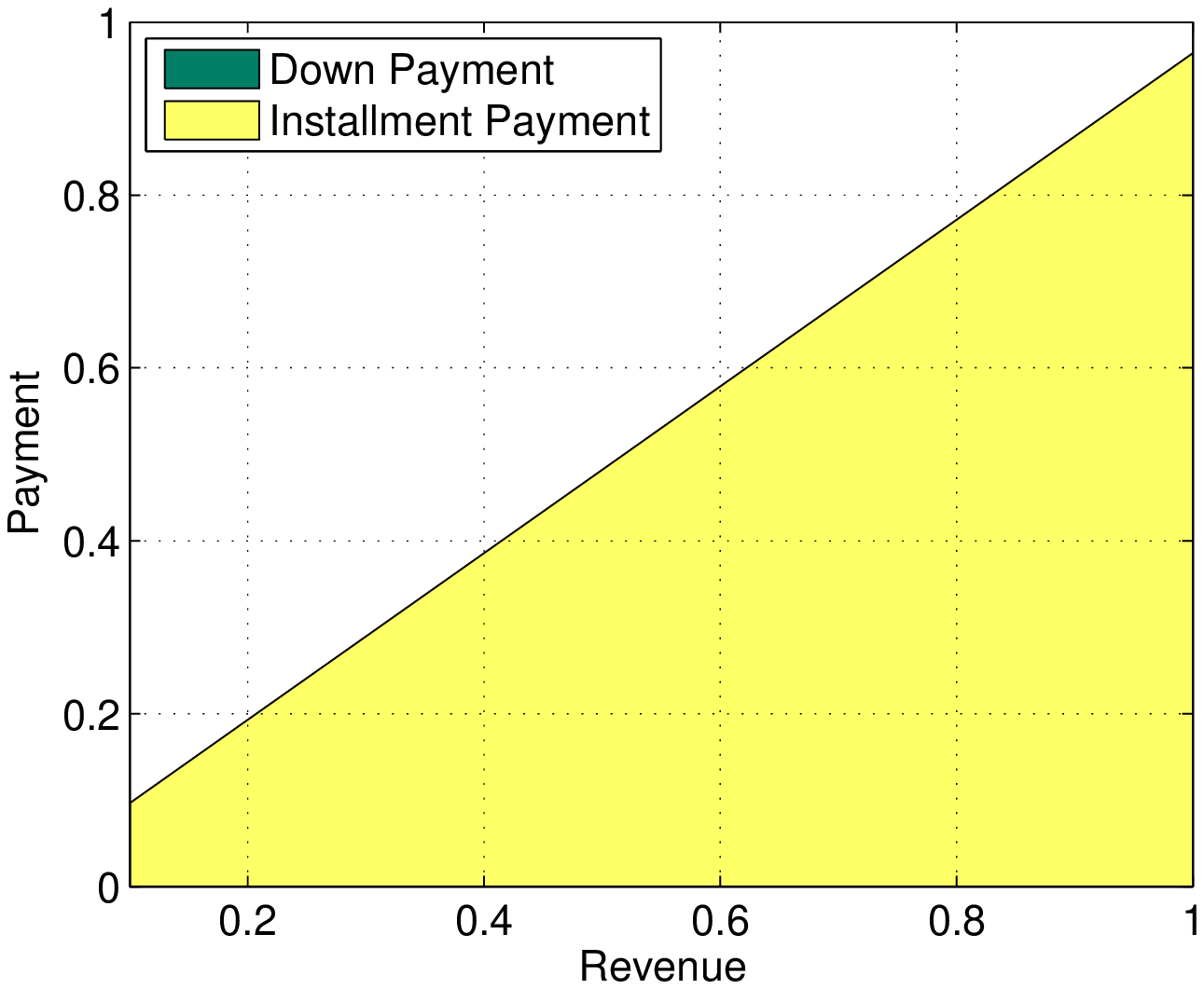}
 \caption{Revenue $R$}
 \label{fig:LowR}
 \end{subfigure}
  \begin{subfigure}[b]{0.32\textwidth}
 \centering
 \includegraphics[width=\columnwidth,height=0.9\textwidth]{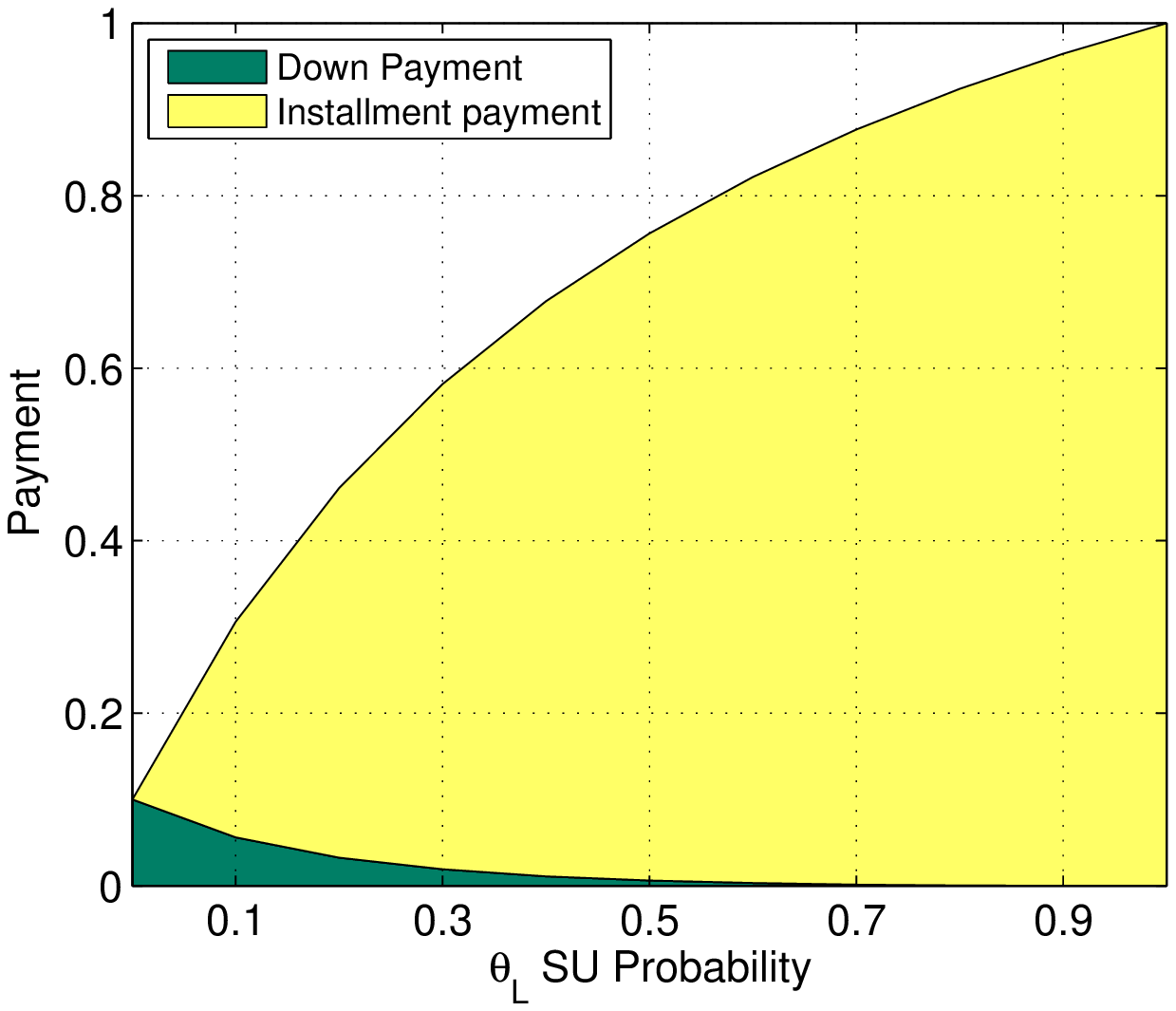}
 \caption{$\theta_L$ SU Probability $\beta$}
 \label{fig:Lowb}
 \end{subfigure}
\centering
\caption{The financing contract for $\theta_L$ SU as parameters vary.}
\label{fig:Low}
\end{figure*}

In Fig. \ref{fig:High}, we show the structure of the financing contract for $\theta_H$ SU when both \emph{adverse selection} and \emph{moral hazard} are present. We see that, with the varying of the three parameters, the installment payment $r_H$ remains $0$, as we have obtained in the previous section. The highest type SU should always pay all the money at the down payment, since cash auction has the highest trading efficiency, but at the cost of PU's revenue.

From Fig. \ref{fig:Highc} we see that, as the cost coefficient $c$ increases, the down payment (i.e., the price of the spectrum) decreases. This result is intuitive in the sense that, when the SU's cost of generating revenue by utilizing the spectrum increases, the SU will be less likely to participate. Thus, the PU must lower its price to attract SU's participation. Otherwise, the vacant spectrum is wasted and zero payoff is obtained by the PU.

In Fig. \ref{fig:HighR} we see that, as the SU's revenue $R$ by ``running'' on the PU's spectrum increases, the cash payment required from the PU increases. This result is also easy to see as if the spectrum can bring more revenue for the SU, the spectrum's value is higher. Thus, the PU would definitely assigned a higher price for the spectrum.

Fig. \ref{fig:Highb} shows when the PU's probability of trading with a $\theta_H$ SU increases, it will also rise the spectrum's price. As we have defined in the system model, the SU's successful probability of obtaining a revenue is $\theta e$. Therefore, under the same effort $e$, the high capable SU will bring a larger expected revenue than low capable SU, as $\theta_H > \theta_L$. Thus, similar to Fig. \ref{fig:HighR}, the PU will raise the price as the value of spectrum increases.

Fig. \ref{fig:Low} is similar to Fig. \ref{fig:High}, as we are showing the financing contract for the $\theta_L$ SU. While different from Fig. \ref{fig:High} is that, the PU asks for installment payment only from the lowest type SU, instead of only down payment when the SU is high capable. This result is intuitive in the sense that, lowest type SU is less possible to bring good outcome. Thus, in order to push the SU to work harder, the PU will ask the money to be paid after the SU has generated revenue from using the spectrum as we have stated in the previous section.

In Fig. \ref{fig:Lowc}, the SU's payment does not affect by the increasing of cost coefficient, since the cost coefficient only affects the down payment. While the results in Fig. \ref{fig:LowR} and Fig. \ref{fig:Lowb} are similar to that in Fig. \ref{fig:High}. That the increasing of revenue and probability will positively affect the payment from the SU. Fig. \ref{fig:Lowb} shows the optimal contract when the PU's probability of trading with a $\theta_L$ SU increases. As the PU becomes more certain that it is trading with a low capable SU with less cash in hand, it will lower the down payment first, but ask for more installment payment instead, which is the SU's price of paying less cash at first.

\begin{figure*}
 \begin{subfigure}[b]{0.32\textwidth}
 \centering
 \includegraphics[width=\columnwidth,height=0.9\textwidth]{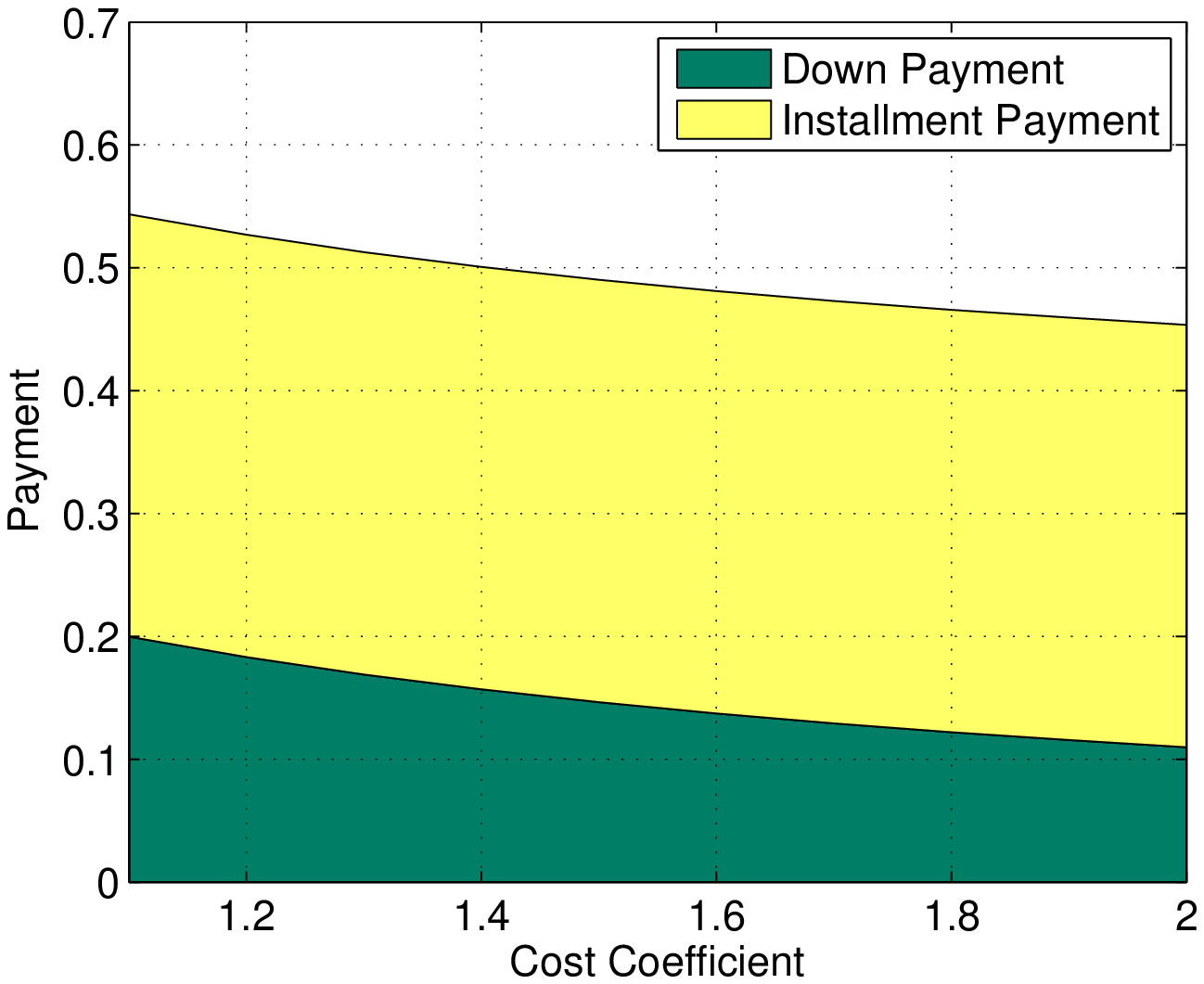}
 \caption{Cost Coefficient $c$}
 \label{fig:midc}
 \end{subfigure}
 \begin{subfigure}[b]{0.32\textwidth}
 \centering
 \includegraphics[width=\columnwidth,height=0.9\textwidth]{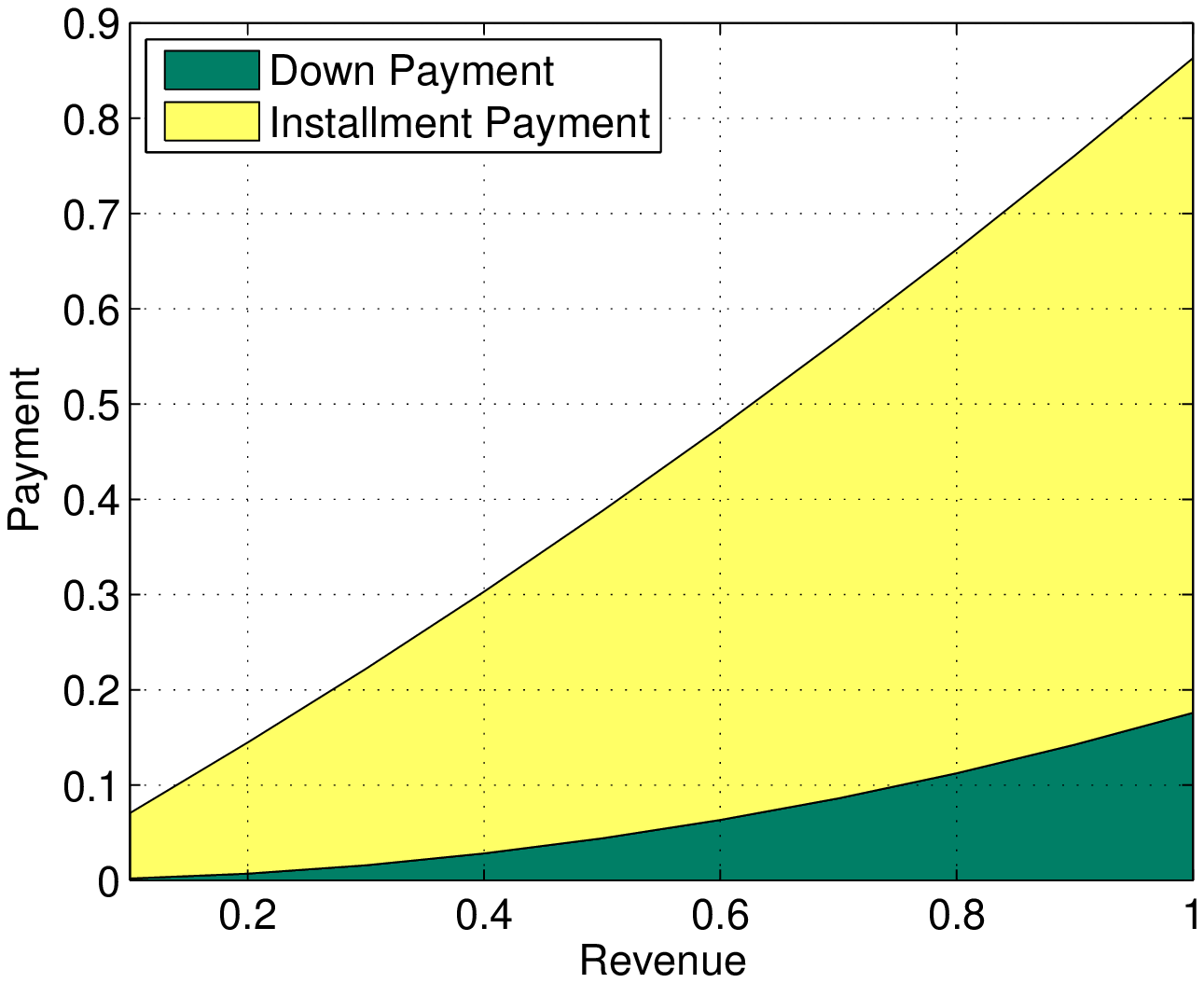}
 \caption{Revenue $R$}
 \label{fig:midR}
 \end{subfigure}
  \begin{subfigure}[b]{0.32\textwidth}
 \centering
 \includegraphics[width=\columnwidth,height=0.9\textwidth]{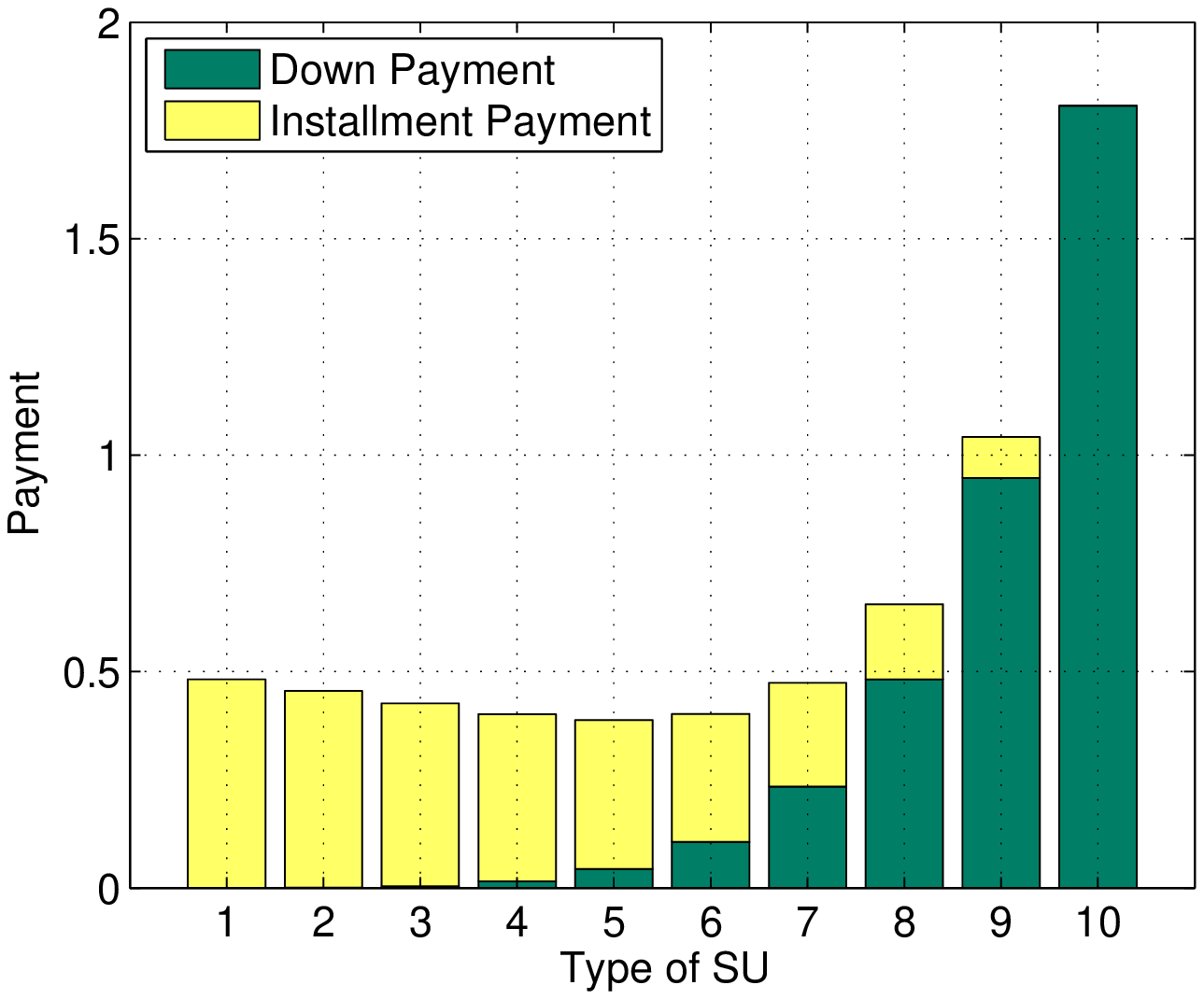}
 \caption{Financing contract for each SU}
 \label{fig:midT}
 \end{subfigure}
\centering
\caption{The financing contract for $\theta_M$ SU as parameters vary.}
\label{fig:middel}
\end{figure*}

\begin{figure*}
 \begin{subfigure}[b]{0.32\textwidth}
 \centering
 \includegraphics[width=\columnwidth,height=0.9\textwidth]{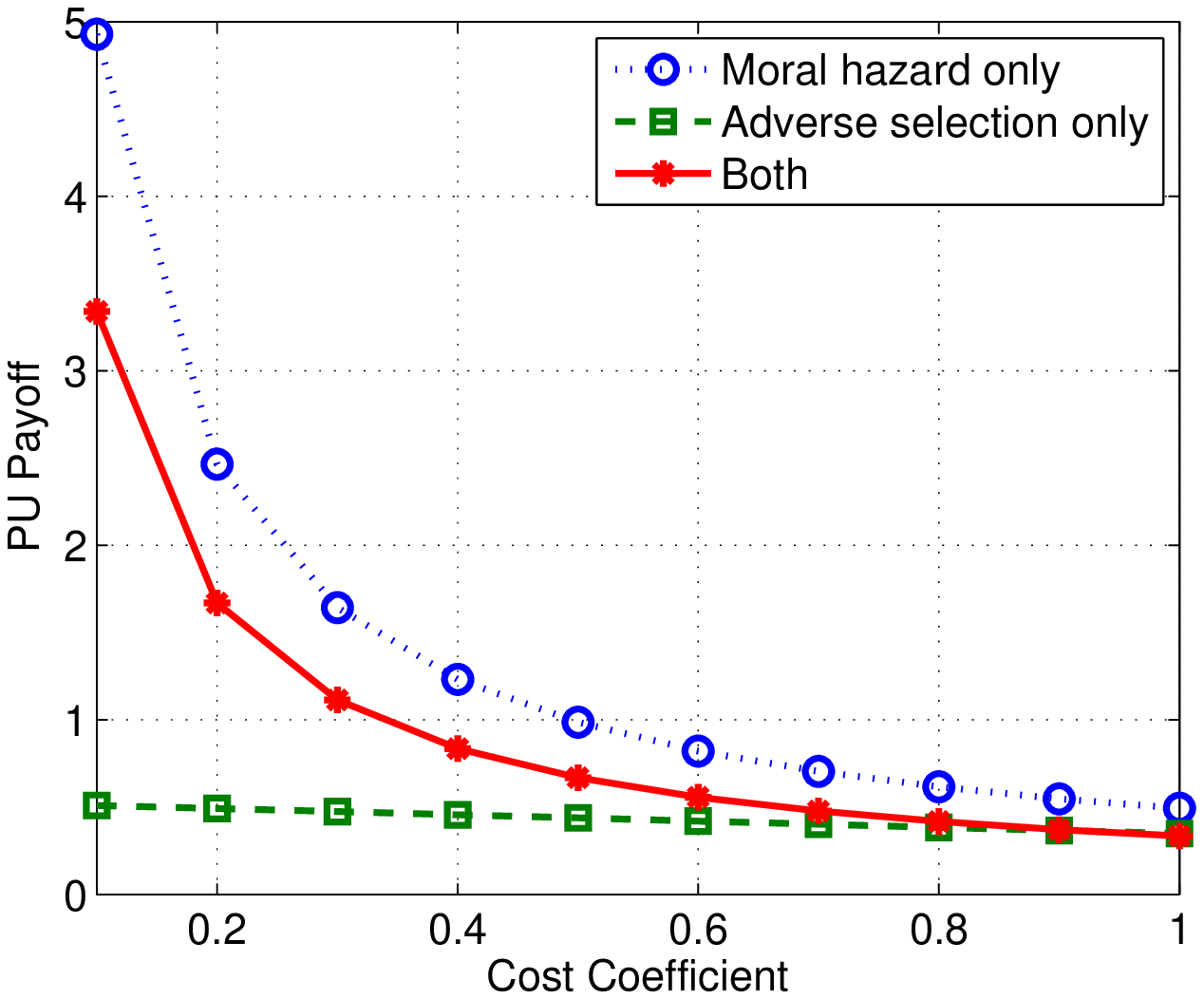}
 \caption{PU's payoff}
 \label{fig:costp}
 \end{subfigure}
 \begin{subfigure}[b]{0.32\textwidth}
 \centering
 \includegraphics[width=\columnwidth,height=0.9\textwidth]{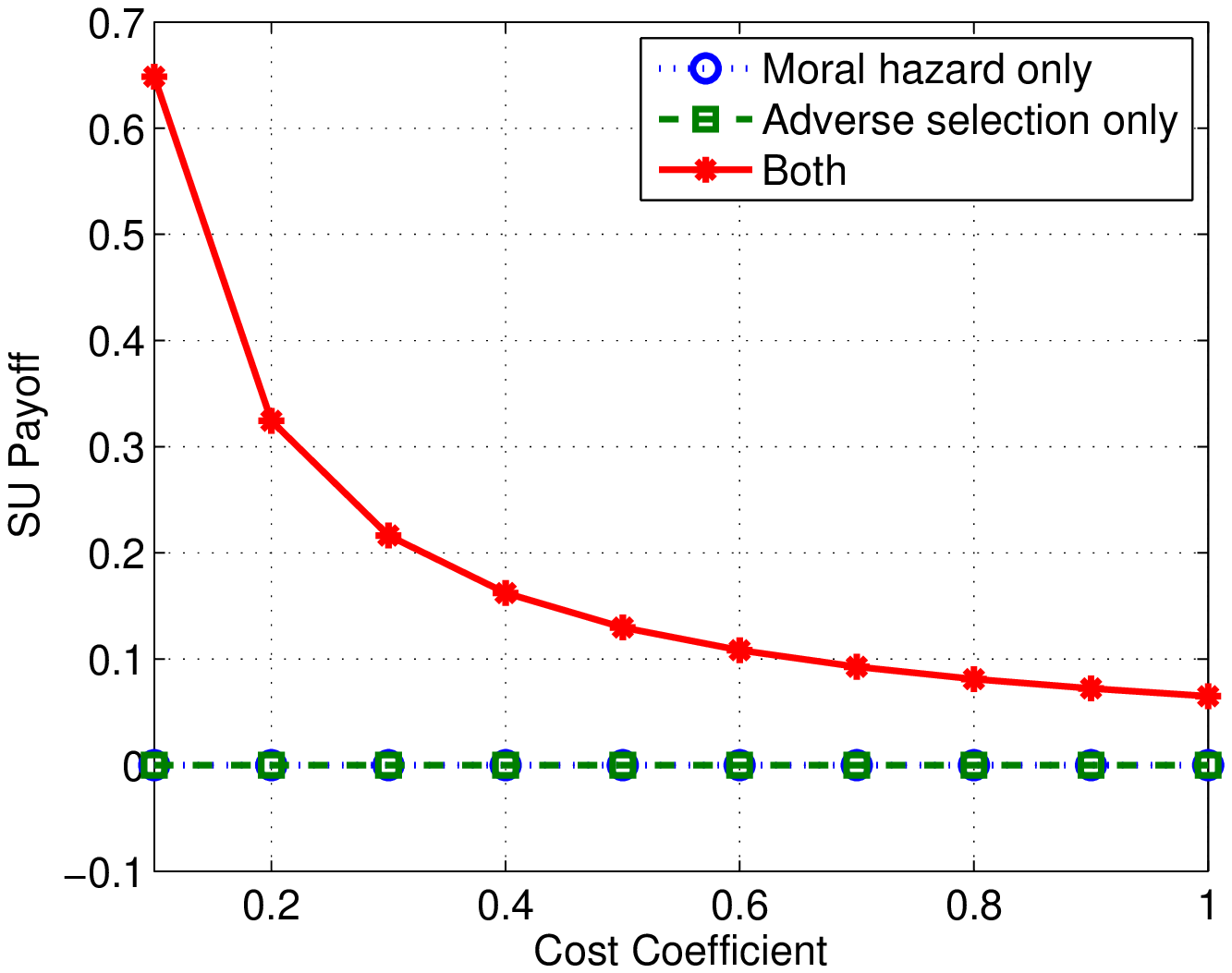}
 \caption{SU's payoff}
 \label{fig:costs}
 \end{subfigure}
  \begin{subfigure}[b]{0.32\textwidth}
 \centering
 \includegraphics[width=\columnwidth,height=0.9\textwidth]{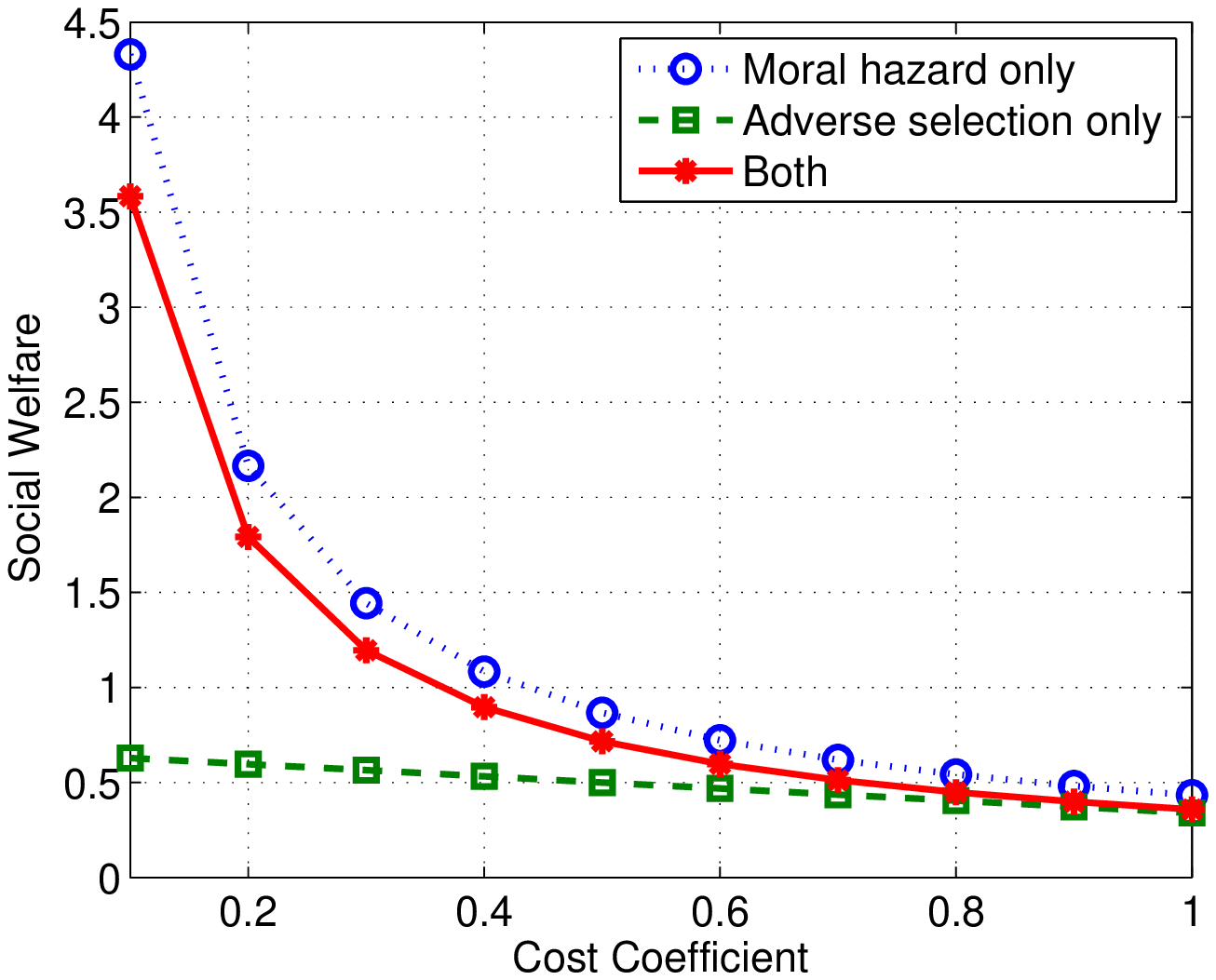}
 \caption{Social welfare}
 \label{fig:costw}
 \end{subfigure}
\centering
\caption{The system performance as the cost coefficient $c$ varies.}
\label{fig:cost}
\end{figure*}

In Fig. \ref{fig:middel} we are showing the financing contract for other types of SU. What different from Fig. \ref{fig:High} and Fig. \ref{fig:Low} is that, the PU asks for both down and installment payment from the SU, instead of only down or installment payment. Since for the rest of SU, the PU will combine the use of down payment and installment payment to incentivize the SU. Similar results are obtained in Fig. \ref{fig:midc} and Fig. \ref{fig:midR} that the overall payment is decreasing with cost coefficient and increasing with revenue.

Furthermore, we are shown the detail of of the financing contract for different types of SU in Fig. \ref{fig:midT} for the case when $n=10$. From type $1$ to type $10$, the down payment is increasing while the installment payment is decreasing. Overall, the highest type SU brings the maximum payment to PU, but the lowest type SU does not necessary to be the one that pays the lowest.

\subsection{System Performance}\label{subsec:SysPer}
\begin{figure*}
 \begin{subfigure}[b]{0.32\textwidth}
 \centering
 \includegraphics[width=\columnwidth,height=0.9\textwidth]{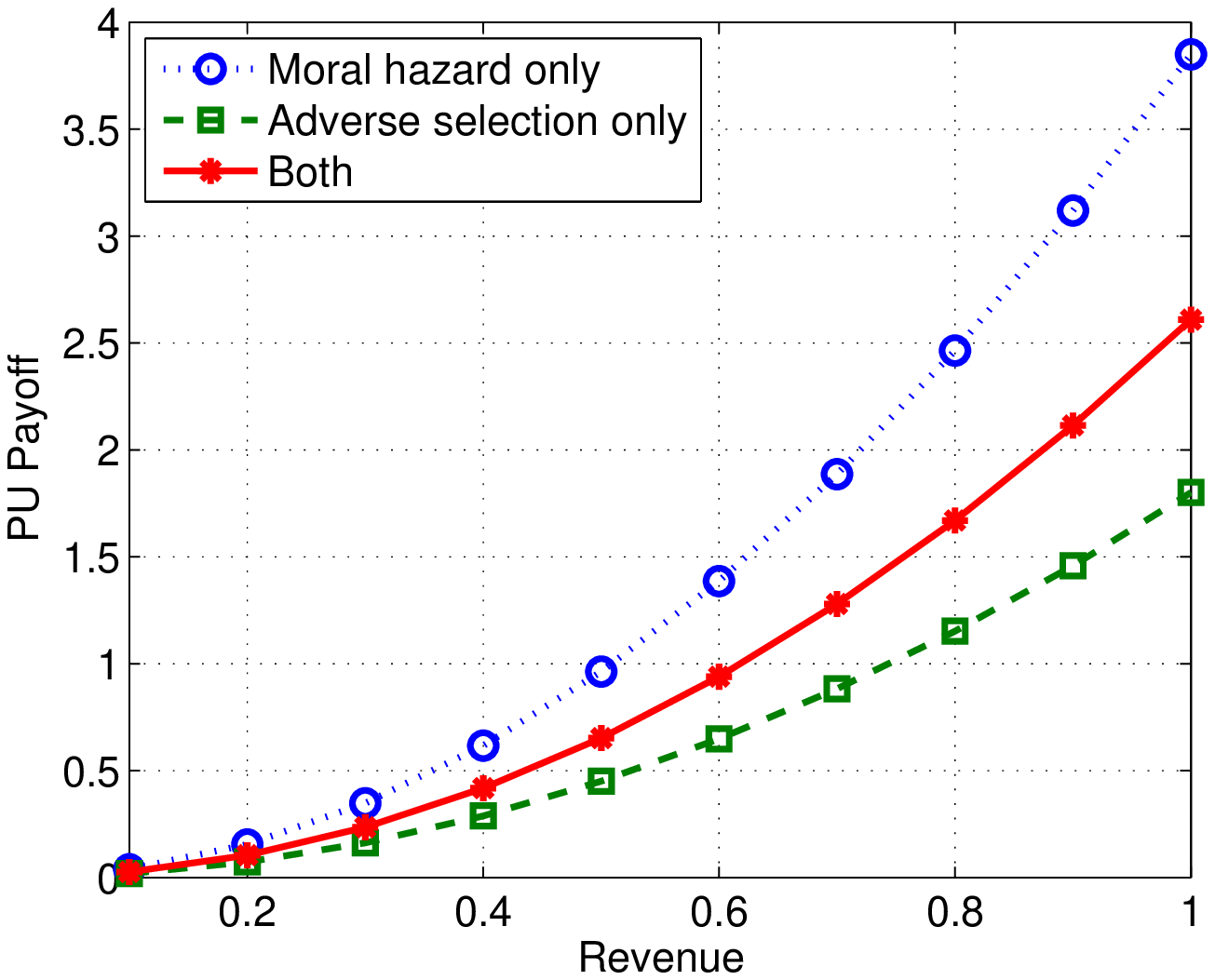}
 \caption{PU's payoff}
 \label{fig:revenuep}
 \end{subfigure}
 \begin{subfigure}[b]{0.32\textwidth}
 \centering
 \includegraphics[width=\columnwidth,height=0.9\textwidth]{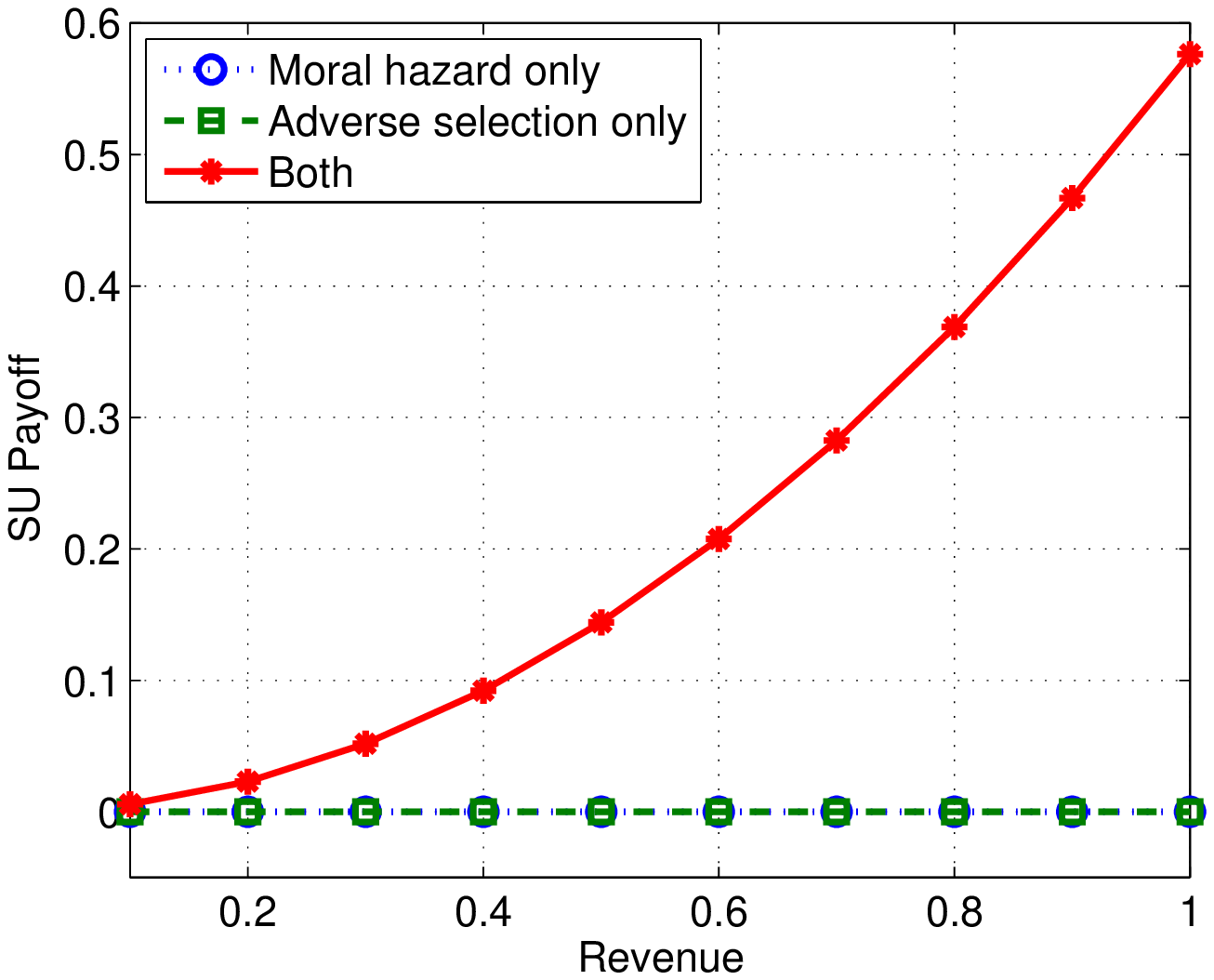}
 \caption{SU's payoff}
 \label{fig:revenues}
 \end{subfigure}
  \begin{subfigure}[b]{0.32\textwidth}
 \centering
 \includegraphics[width=\columnwidth,height=0.9\textwidth]{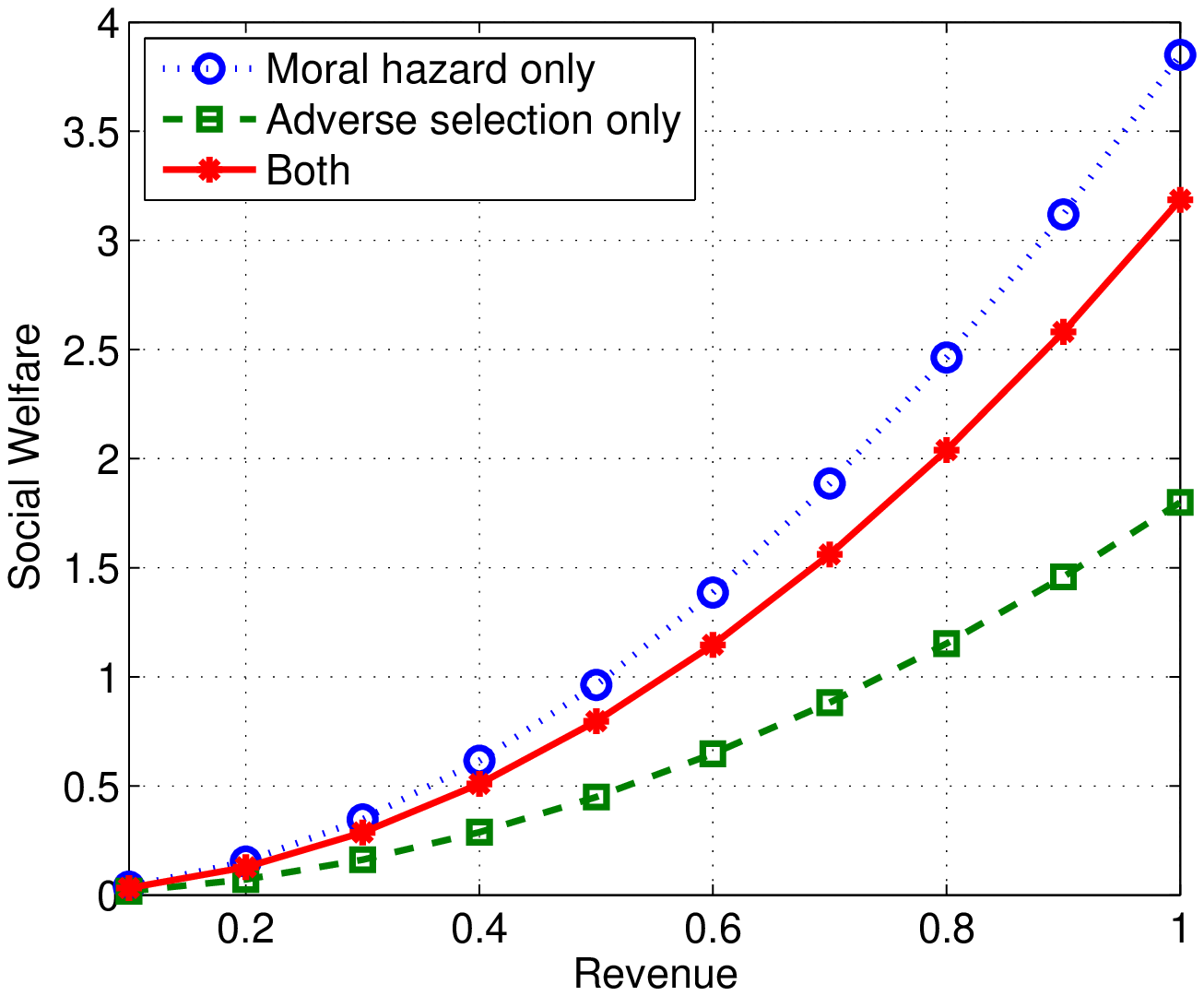}
 \caption{Social welfare}
 \label{fig:revenuew}
 \end{subfigure}
\centering
\caption{The system performance as the revenue $R$ varies.}
\label{fig:revenue}
\end{figure*}
\begin{figure*}
 \begin{subfigure}[b]{0.32\textwidth}
 \centering
 \includegraphics[width=\columnwidth,height=0.9\textwidth]{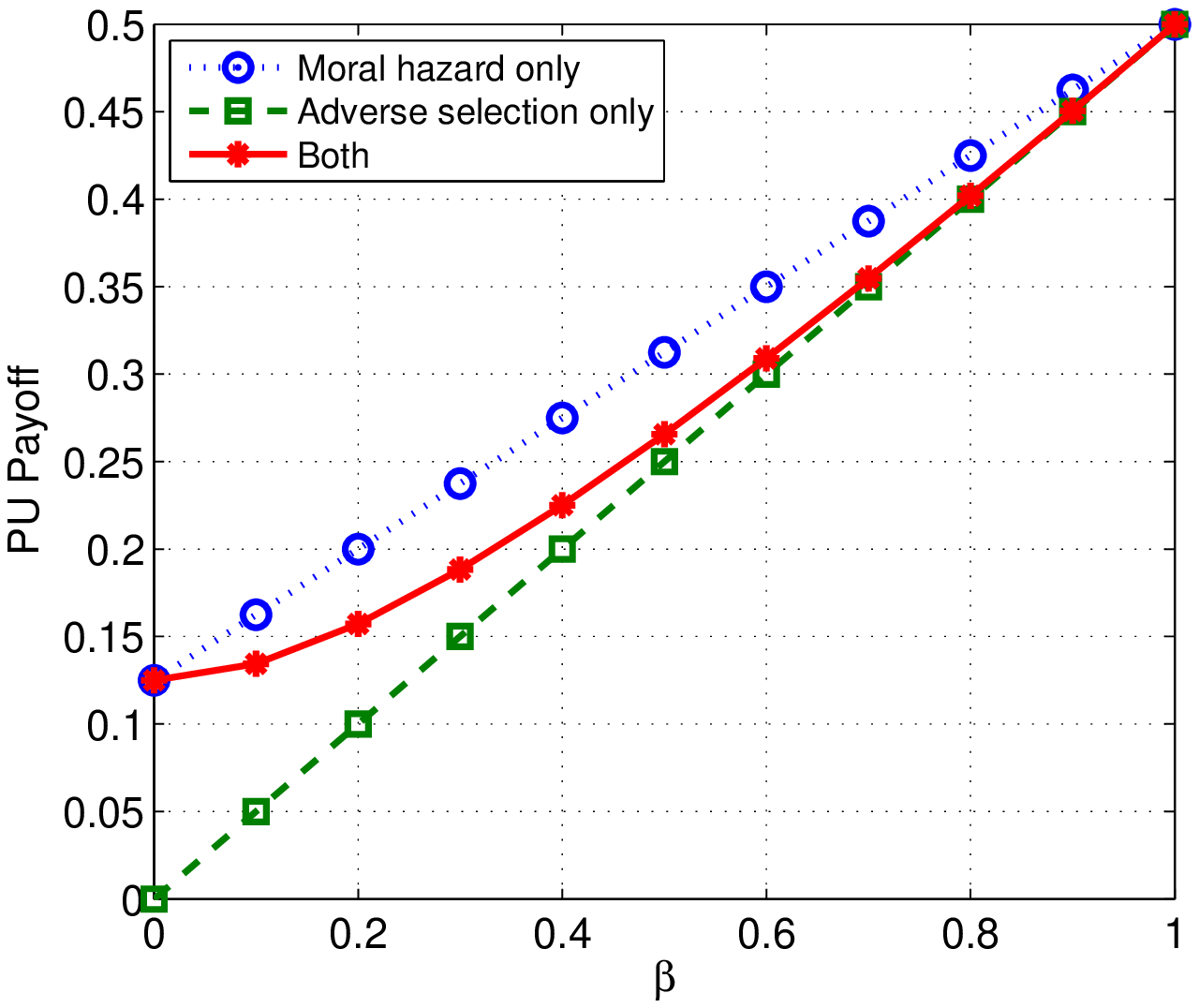}
 \caption{PU's payoff}
 \label{fig:distp}
 \end{subfigure}
 \begin{subfigure}[b]{0.32\textwidth}
 \centering
 \includegraphics[width=\columnwidth,height=0.9\textwidth]{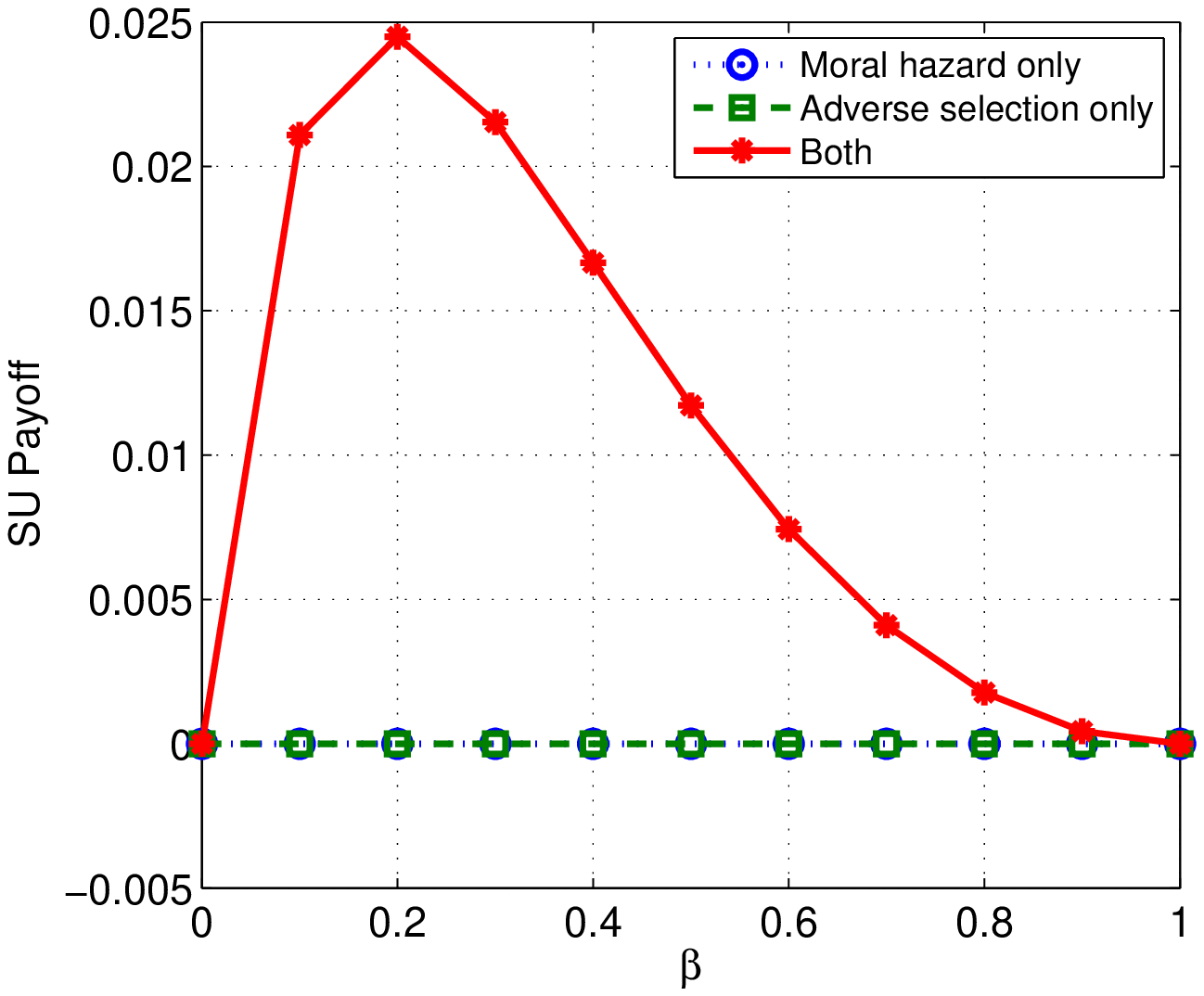}
 \caption{SU's payoff}
 \label{fig:dists}
 \end{subfigure}
  \begin{subfigure}[b]{0.32\textwidth}
 \centering
 \includegraphics[width=\columnwidth,height=0.9\textwidth]{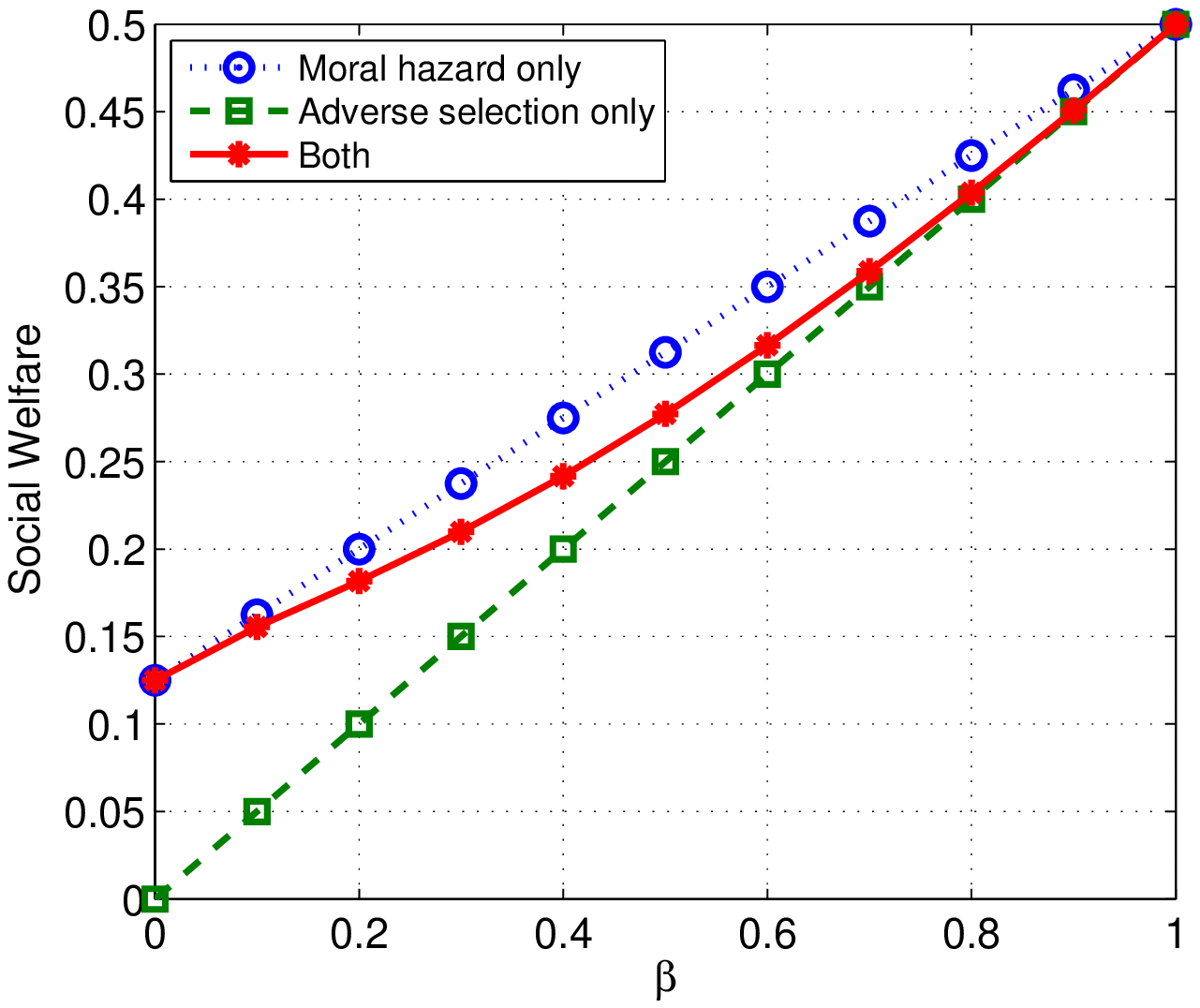}
 \caption{Social welfare}
 \label{fig:distw}
 \end{subfigure}
\centering
\caption{The system performance as the $\theta_H$ SU probability $\beta$ varies.}
\label{fig:dist}
\end{figure*}

From Fig. \ref{fig:cost} to Fig. \ref{fig:dist}, we compare the system performance under the three scenarios we have proposed: \emph{moral hazard} only, \emph{adverse selection} only, and when both are present. The \emph{adverse selection} only and \emph{moral hazard} only also represent the one-shot cash auction and linear pricing strategy in previous works, respectively. In the following part, we will give a detailed analysis of the cost coefficient $c$, revenue $R$, and distribution $\beta$'s effects on the system performance.

\subsubsection{Cost Coefficient}\label{subsec:Analy}
In Fig. \ref{fig:cost}, we vary the value of the cost coefficient $c$ to see the effects on the PU's and SU's payoffs, and the social welfare in the three scenarios. As we can see, PU's and SU's payoffs and social welfare decrease as the cost coefficient increases, except the SU's payoff under \emph{moral hazard} only and \emph{adverse selection} only scenarios. Under those two extreme cases, the PU has the full acknowledgment of either the SU's cash in hand or the effort put into using the spectrum. Thus, the PU can extract as much revenue as possible from the SU, which leaves the SU with zero payoff. The reason for the decreasing of payoffs and social welfare is similar to the analysis we gave for Fig. \ref{fig:Highc} and Fig. \ref{fig:Lowc} that as the cost increasing, the price for the spectrum will decrease to attract the SU. As a result, the payoffs of the PU and SU, together with social welfare, will decrease.

\subsubsection{Revenue}\label{subsec:Comp}
In Fig. \ref{fig:revenue}, we try to see the PU and SU's payoffs, and the social welfare when the revenue $R$ can be generated from using the spectrum increases. We see that the payoffs and social welfare increase with the revenue except the SU's payoff under \emph{moral hazard} only and \emph{adverse selection} only scenarios. The increase of payoffs and social welfare with the revenue $R$ is easy to understand as
we have explained in the previous paragraph that the PU will extract all the information rent from the SU.

\subsubsection{Distribution}\label{subsubsec:Indep}
In Fig. \ref{fig:dist}, we see that PU's payoff and the social welfare increase as $\beta$ gets larger. The reason for this result is the same as we have explained for Fig. \ref{fig:Highb} and Fig. \ref{fig:Lowb}, as the PU will ask for more money if it believes that it is facing a high capable SU. However, the increase of $\beta$ has a negative effect on the SU's payoff as the PU is trying to extract revenue from the SU.

Overall, from Figs. \ref{fig:cost}-\ref{fig:dist}, we see that, the two extreme cases serve as the upper and lower bounds, respectively. The PU's payoff in the general case where both \emph{moral hazard} and \emph{adverse selection} present lies between the two extreme cases.
\section{Conclusions}\label{sec:Conclusion}

In this paper, we have proposed a financing contract to address the problem of non-cash auction for spectrum trading in a cognitive radio network. The problem in both discrete and continuous cases have been formulated by considering a joint \emph{adverse selection} and \emph{moral hazard} model of the second user. In addition, we have solved and analysed the problems under three different scenarios, i.e., two extreme cases where only \emph{adverse selection} or \emph{moral hazard} is present, and the general case where both are present. The \emph{adverse selection}'s and \emph{moral hazard}'s affects on PU's and SU's payoffs have been discussed, as well as their connections with previous works. Through extensive derivations and simulations, different parameters' effects on the system performance have been shown. Comparisons among the three proposed mechanisms have been shown, which are also serving as the comparisons with previous works. The simulation results have also proved the previous analysis about the financing contract for all considered scenarios, that 1) the higher the type of SU, the more overall payment is required from the PU; 2) the PU's expected revenue is decreasing with cost coefficient, risk averse degree, and probability of meeting a low type SU; 3) non-cash auction will bring higher expected revenue than full-payment required mechanism, but lower than the mechanism that requires no down payment.

\end{spacing}
\begin{spacing}{1}
\bibliographystyle{IEEEtran}

\bibliography{./AdMh_Single}
\end{spacing}
\end{document}